\documentclass[3p]{elsarticle}

\usepackage{setspace}
\usepackage{amsmath, amsthm, amssymb, mathrsfs}
\usepackage{bbm}
\usepackage{booktabs}
\usepackage{graphicx}
\usepackage[hidelinks]{hyperref}
\usepackage{pdflscape}

\makeatletter
\def\ps@pprintTitle{%
   \let\@oddhead\@empty
   \let\@evenhead\@empty
   \let\@oddfoot\@empty
   \let\@evenfoot\@oddfoot
}
\makeatother

\newpageafter{abstract}

\def\sym#1{\ifmmode^{#1}\else\(^{#1}\)\fi}

\begin{document}

\begin{frontmatter}

\title{Interdisciplinary research and technological impact: Evidence from biomedicine}

\author{Qing Ke\corref{corrauthor}}
\address{School of Data Science, City University of Hong Kong, Hong Kong, China}
\cortext[corrauthor]{Corresponding author}
\ead{q.ke@cityu.edu.hk}

\begin{abstract}
Interdisciplinary research (IDR) has been considered as an important source for scientific breakthroughs and as a solution to today's complex societal challenges. While ample empirical evidence has suggested its benefits within the academia such as better creativity and higher scientific impact and visibility, its societal benefits---a key argument originally used for promoting IDR---remain relatively unexplored. Here, we study one aspect of societal benefits, that is contributing to the development of patented technologies, and examine how IDR papers are referenced as ``prior art'' by patents over time. We draw on a large sample of biomedical papers published in 23 years and measure the degree of interdisciplinarity of a paper using three popular indicators, namely variety, balance, and disparity. We find that papers that cites more fields (variety) and whose distributions over those cited fields are more even (balance) are more likely to receive patent citations, but both effects can be offset if papers draw upon more distant fields (disparity). These associations are consistent across different citation-window lengths. We further find that conditional on receiving patent citations, the intensity of their technological impact, as measured as both raw and quality-adjusted number of citing patents, increases with balance and disparity. Our work may have policy implications for interdisciplinary research and scientific and technological impact.
\end{abstract}


\end{frontmatter}

\section{Introduction}

\citet{NAP11153} defined interdisciplinary research (IDR) as research that ``integrates ... from two or more disciplines or bodies of specialized knowledge to advance fundamental understanding or to solve problems whose solutions are beyond the scope of a single discipline or area of research practice.'' IDR has received intense attentions from diverse stakeholders in science. For example, science policymakers have been constantly discussing IDR \citep{Wagner-approache-2011}; \citet{Metzger-1999-idr} described it as ``a `mantra' of contemporary science policy'' \citep{feller2006multiple}. Funding bodies have been actively promoting interdisciplinary workings and encouraging IDR proposals \citep{Lowe-reflexive-2006, Millar2013idr}. Academic institutions have created interdisciplinary training programs \citep{Misra2009evaluating} and established cross-department interdisciplinary centers to better foster IDR \citep{bordons1999measuring, Cech-nurturing-2004, Biancani2014semiformal, hackett2021synthesis}.

These efforts to supporting IDR rest on the premise that it is immensely beneficial in many aspects. IDR is expected among scholars and science policymakers to have the potential to be a key to science advancement and to produce breakthrough contributions in science \citep{Jacobs-2009-idr, Porter2006idr, szostak2008classification}. Such an expectation is grounded on the proposition that IDR makes novel combinations of existing knowledge, which makes it more likely to generate high scientific impact \citep{Uzzi-atypical-2013}. In this perspective, IDR has been considered as a solution to today's complex societal challenges such as climate change and sustainability, as solving them requires knowledge that transcends traditional disciplinary boundary \citep{Ledford-how-2015}.

Substantial empirical evidence has indeed suggested that IDR bring forth benefits. These include better creativity in research \citep{Heinze-2009-org, Jacobs-2009-idr}, higher scientific impact for IDR papers \citep{Steele-impact-2000}, better academic impact, visibility, and long-term funding performance for scientists involved in IDR \citep{Leahey-prominent-2017, sun2021interdisciplinary}, and higher likelihood to obtain academic positions for doctoral graduates who conduct interdisciplinary dissertation research \citep{Millar2013idr}. However, these are benefits confined to the academia, and whether IDR has societal benefits---one of the key arguments originally used for promoting IDR---remains comparatively unknown.

The purpose of this study, therefore, is to expand empirical characterizations of the benefits of IDR from the dominantly studied academic benefits to the understanding of societal benefits of IDR. We focus on the technological domain and examine whether IDR can be used in the development of patented technologies. We present the first, to our best knowledge, bibliometric study that explores the relationship between the extent of IDR of papers and their technological impact, by tracing their received citations made by patents. Our analyses rely on three large-scale datasets: (1) over 5.4 million papers in the biomedicine domain; (2) 6.5 million USPTO (U.S. Patent and Trademark Office) utility patents; and (3) 2.6 million citations from patents to papers. In so doing, technological development is represented by patents, following existing literature in innovation studies \citep{Fleming-science-2004}, and the contribution of science to technology is captured by patent-to-paper citations, as they signal and better capture knowledge spillovers than patent-to-patent citations do \citep{Roach-lens-2013}. Given the lack of consensus on how to measure IDR and to capture different dimensions of IDR, we use three distinct indicators, namely variety, balance, and disparity, as well as an integrated metric called Rao-Stirling index, to quantify the degree of interdisciplinarity of a paper. We find that the effects of IDR on technological impact is dependent on different dimensions. In particular, variety has a positive, but small, effect on the likelihood of getting patent citations. Balance has a positive, sizable effect. Disparity, on the other hand, has a negative effect. These associations are robust even after controlling for several confounders like paper quality and journal Impact Factor and consistent across different citation-window lengths. In our further analyses that focus on papers with patent citations, we find that the intensity of technological impact, as measured as both raw and quality-adjusted number of citing patents, increases with both balance and disparity.

\section{Background literature} \label{sec:lit}

\subsection{Interdisciplinary research}

What is interdisciplinary research? According to \citet{NAP11153}, IDR is a mode of research that integrates knowledge from multiple disciplines for fundamental understanding or problem solving. Distinct from multidisciplinary research, which is a juxtaposition of disciplines, IDR emphasizes the integration of knowledge, methods or theories from different disciplines. IDR has experienced a surge of interests in the last decades. It has been increasingly prevalent across both natural and social sciences: Academic institutions initialize the creation of cross-department research centers to foster IDR \citep{Biancani2014semiformal, bordons1999measuring, Cech-nurturing-2004}; interdisciplinary training programmes such as the Integrative Graduate Education and Research Traineeship have been developed; and funding agencies have been encouraging and promoting interdisciplinarity in grant proposals. These burgeoning initiatives have motivated widespread academic interests in IDR in the fields of sociology of science, innovation studies, and science policy \citep{Jacobs-2009-idr, Leahey-prominent-2017, Este-relation-2019}.

Underlying the enthusiasms about IDR is the shared expectation that IDR is more likely to give rise to scientific breakthroughs and innovations and to solve complex societal problems. First, IDR integrates bodies of knowledge that span different disciplines, and the theory of knowledge recombination implicates that such a boundary-spanning search results in scientific findings that make atypical combinations of knowledge, which are more likely to have high scientific impact \citep{Uzzi-atypical-2013}. This recombinant search has also been suggested as the primary mode for the production of innovations \citep{Lee2001recombinant}. Second, today's societal problems, such as climate change, are rarely disciplinary ones, and their solutions are particularly suited to IDR, as it encompasses perspectives and approaches from a plurality of disciplines \citep{Ledford-how-2015}.

Apart from these contentions, a large stream of inquiry in the literature has been the proposals of indicators for IDR and subsequent examinations of its relationship with scientific impact. Broadly speaking, these studies can be viewed as empirical understandings of the benefits of IDR. They all build IDR measures using discipline information of cited references but differ in the ways in which IDR metrics are constructed. 

Early versions of IDR indicators are one dimensional. \citet{Rinia-influence-2001} defined interdisciplinary papers as those published in journals whose disciplines are different from the main focal program of interest (physics) and found no evidence of bibliometric or peer-review bias against IDR. Similarly, \citet{Lariviere-relation-2010} defined interdisciplinarity of a paper as the fraction of its cited references that were published in journals of other disciplines. \citet{Levitt-multi-2008} looked at papers published in single-subject journals vs papers in multiple-subject ones. They found similar citations for the two sets of papers in social sciences, but for life, health, and physical sciences, citations of papers in the former category are larger than that of papers in the latter group. Another class of IDR indicators considers the diversity of cited disciplines. \citet{Steele-impact-2000} quantified IDR using Brillouin's diversity index and found a positive association with citation rate. Other works used diversity measures to quantify interdisciplinarity of journals \citep{Silva-quantify-2013} and authors \citep{Carayol-why-2005}.

Recent development has highlighted that IDR is not only about how diverse cited disciplines are but also about how they are related to each other. This has spurred proposals of multidimensional indicators emphasizing different aspects of IDR. In particular, \citet{stirling2007general} introduced three dimensions of IDR, which are variety, balance, and disparity. Variety counts the number of cited disciplines, balance quantifies the diversity of these disciplines, and disparity measures their relatedness. \citet{Wang-impact-2015} similarly measured IDR through the three facets. \citet{Wang-impact-2015} used the three dimensions of IDR to establish their distinct effects on scientific impact. That is, variety and disparity are negatively linked to short-term citations but positively associated with long-term citations; balance lacks a significant effect on short-term citations, and it is negatively associated with long-term citations. \citet{Yegros-interdisc-2015} presented a similar analysis. Yet, the effects of the three aspects of IDR on scientific impact are not entirely the same as the effects identified in \citet{Wang-impact-2015}. Specifically, \citet{Yegros-interdisc-2015} found that while variety has a positive effect on scientific effect, balance and disparity have negative effects. The literature has also proposed integrated measures of the three indicators, such as the Rao-Stirling (RS) index \citep{Porter2006idr, porter2007measuring} and its variations \citep{Leydesdorff-idr-2019}. \citet{Porter-science-2009} applied the RS index to papers published in 1975--2005 and observed a modest increase of interdisciplinarity. \citet{Cassi-analysing-2017} applied the index to institutions. Recently, \citet{Gates-nature-2019} performed a large-scale analysis of 19 million articles from 1900 to 2017 and observed an increasing interdisciplinarity across disciplines. Motivated by the question of whether synthesis centers---scientific organizations catalyzing and supporting integration research---actually synthesize, \citet{hackett2021synthesis} based Latent Dirichlet Allocation to quantify topical diversity, from the perspectives of both aggregated diversity and particular aspects of diversity (\emph{i.e.}, variety, evenness, and balance). They found that papers from synthesis centers have greater topical variety and evenness, yet less disparity, than do reference papers.

In addition to these publication-level analyses, some other extant works have instead looked at scientists. \citet{Leahey-prominent-2017} found that scientists involved in IDR have more academic impact and visibility, although they are less productive. \citet{Millar2013idr} found that doctoral students whose dissertation research is more interdisciplinary are more likely to secure academic positions. Focusing on UK research grants, \citet{sun2021interdisciplinary} showed that interdisciplinary researchers achieve better long-term funding performance than their matched, specialized counterparts, despite lower short-term impact for their papers. \citet{fontana2022interdisciplinarity} examined the potential tension between researchers' private and public interests. Using citations as a proxy for a researcher's private interest to build a reputation and the circulation of her papers beyond disciplinary boundaries as a proxy for her public interest for solutions to societal issues, their case studies of researchers affiliated with the University of Florida indicated differing effects of the diverse aspects of IDR (variety, balance, and disparity) on a researcher's reputation and contribution to societal research, confirming that there is a trade-off between private and public interest.

Conducting IDR brings benefits, but not without costs. In particular, modern science has established a discipline-based organization, and each discipline has been developing and in favor its own norms, concepts, objectives, methods, etc. This on one hand provides an intellectual ``safe harbor'' for mono-disciplinary researchers, but at same time puts IDR in a disadvantage in the recognition and appreciation of its value. Subsequently, this makes IDR researchers face numerous difficulties in their careers, such as poor career advancement, low esteem by colleagues, and difficulty in publishing in prestigious venues and in securing grants \citep{Bromham-interdisc-2016, Bruce2004idr, Porter-peer-1985, Klein2008, Rafols-how-2012}.

\subsection{Technological impact of scientific research}

Our discussions thus far have been restricted to costs and benefits of IDR within the academia. IDR has also been widely viewed to possess the potential to contribute to satisfy societal needs. Empirical scrutiny on this contention, however, has been very limited. The goal of this work is to present one aspect of societal benefits of IDR---contributing to the development of patented technologies. Before discussing this relationship, we first briefly survey the role of scientific research in general in the development of technologies.

Scientific research has long been shown to have impact over broader domains beyond the science itself, and one of the most studied such domains is the technological one. The questions of how and through which channels scientific research contributes to technological development have been extensively explored in the innovation studies literature and have become increasingly relevant in science policy discussions. In answering them, the literature has predominantly relied on patent-to-paper citations, assuming that citing a scientific paper in a patent signals that the patent builds upon knowledge presented in the paper and therefore there is knowledge spillovers from the paper to the patent. Early works in this area were from Narin and his colleagues \citep{Narin-status-1992, Narin-linkage-1997, Narin1998linkage}. Their studies pointed out a growing number of citations from US patents to public science and that for the biotechnology industry, this type of citations is more common than other industries. A great attention has been paid to elucidate the factors facilitating science-technology knowledge spillovers, including geography \citep{Belenzon-spreading-2013, Bikard-bridging-2019}, organization type \citep{Bikard-made-2018}, and intrinsic characteristics of scientific knowledge \citep{Ke-tech-2020}, among many others.

A few studies examined to what extent patent-to-paper citations can represent knowledge spillovers. \citet{Meyer-does-2000} used nano-technology patents as a case study and suggested that papers cited in front-pages of patent documents provided background information. \citet{tussen2000technological} focused on Dutch science and found supporting evidence that patent-to-paper citations ``reflect genuine links between science and technology''.

\subsection{Technological impact of IDR}

Turning to whether and how IDR may achieve technological and other types of societal-relevant benefits, although empirical studies are still lacking, there are some discussions about the relationship between IDR and societal impact. \citet{Gallart-relation-2014}, for example, identified that the modality of ``long-range'' IDR---research with the integration of cognitively distant disciplines---is more likely to link to societal impact than other types of IDR. \citet{Este-relation-2019} argued that scientists who have conducted IDR are more likely to involve in the engagement of university-industry interaction. The line of reasoning is that IDR involves cooperation of actors from different disciplines or types of organizations, which makes it difficult to align research goals, approaches, risks, etc. Therefore, there are significant coordination costs arose from collaborations, and scientists who have successfully involved in IDR, as evidenced from publishing IDR papers, have developed necessary cognitive and social skills to reduce the coordination costs. These skills are particularly useful when interacting with diverse types of partners including industry ones, therefore making them more likely to engage in university-industry interaction. \citet{Este-relation-2019} further tested empirically the relationship between scientists' IDR orientation and their engagement in university-industry interaction, finding that IDR has an impact for all the four examined modes of engagement (academic entrepreneurship, technology transfer, research partnership, and research services). 

Following the line of university-industry interaction, \citet{Giuliani2010who} studied how researcher's characteristics and institutional specificities may explain the propensity to engage in different types of university-industry linkages and found that researcher's age, gender, and centrality in the academic system is more important than publications or formal degrees. \citet{Rijnsoever2008resource} viewed networks as resources for university researchers' competitive advantages for career development and showed that networks formed with other university researchers were helpful for careers, but university-industry networks were not. They further found that although there were declines in a researcher's collaborations within the academia, collaborations with industry continued to increase throughout a researcher's career.

\section{Data and methods} \label{sec:data}

\subsection{Biomedical paper sample selection}

As our interest resides in the biomedicine area, we use MEDLINE as our primary source for publication data. MEDLINE is a widely used database for biomedical research literature that contains more than 28 million documents.\footnote{\url{https://www.nlm.nih.gov/medline/medline_overview.html}} We download the data from \url {https://www.nlm.nih.gov/databases/download/pubmed_medline.html} and select documents published between 1980 and 2002. We focus on this period because we later need to track how these documents get cited by patents that are granted from 1976 to 2012. Therefore, there are at least 10 years for these MEDLINE documents to accumulate patent citations.

Next, we apply three filters to arrive at our final corpus of papers. First, we filter out documents that are not research articles, as MEDLINE indexes myriad types of documents, such as ``journal article'', ``biography'', etc. Here we adopt the operationalization of research articles used by \texttt{iCite}, a bibliometric tool developed by NIH, which categorizes a document as a research article if its ``publication type'' tags in MEDLINE contain at least one pre-defined qualifying tag but do not have any pre-defined disqualifying tags.\footnote{The lists of qualifying and disqualifying tags can be found at \url{https://icite.od.nih.gov/user_guide?page_id=ug_data}} Second, using the crosswalk file between PubMed ID (PMID), an identifier for MEDLINE documents, and Web of Science (WoS) accession number provided by WoS, we match MEDLINE documents with WoS database and obtain additional bibliographical information from the WoS database. We then filter out papers that are not ``article'', ``letter'',  or ``note'' based on their ``document type'' in the WoS. Third, we filter out papers whose research fields, as specified as WoS Subject Category (SC), are not directly related to biomedical research---mostly social sciences and humanities (SSH) fields, as papers from those fields may be less likely to get patent citations. The reason that MEDLINE contains SSH papers is because MEDLINE also indexes research from these areas that pertains to biomedicine.

\begin{table*}[t!]
\centering
\caption{Number and percentage of papers by field, as well as percentage of papers that get cited by patents within 5, 10, and 15 years after publication. Papers that belong to multiple fields are counted multiple times.\label{tab:sc-cnt}}
\begin{tabular}{l r c | r r r}
\toprule
                                                &         &      & \multicolumn{3}{c}{\% cited by patents after} \\
                                         Field  &  Papers &  \%  & \multicolumn{1}{r}{5 y.} & \multicolumn{1}{r}{10 y.} & \multicolumn{1}{r}{15 y.} \\
\midrule
              Biochemistry \& Molecular Biology &  651037 & 7.91 &            5.49 &            14.29 &            18.37 \\
                       Pharmacology \& Pharmacy &  359079 & 4.36 &            2.70 &             6.84 &             9.22 \\
                  Medicine, General \& Internal &  352475 & 4.28 &            0.78 &             1.84 &             2.49 \\
                                        Surgery &  337538 & 4.10 &            1.24 &             2.96 &             3.94 \\
                                  Neurosciences &  328591 & 3.99 &            1.66 &             4.66 &             6.29 \\
                                     Immunology &  282170 & 3.43 &            4.30 &            11.73 &            15.42 \\
                                   Cell Biology &  253703 & 3.08 &            5.05 &            13.43 &            17.13 \\
                                       Oncology &  245098 & 2.98 &            3.36 &             9.07 &            11.87 \\
                             Clinical Neurology &  180076 & 2.19 &            1.13 &             3.06 &             4.31 \\
              Cardiac \& Cardiovascular Systems &  176064 & 2.14 &            2.21 &             5.23 &             6.76 \\
                                     Biophysics &  172271 & 2.09 &            3.54 &             9.39 &            12.44 \\
                           Genetics \& Heredity &  170625 & 2.07 &            4.42 &            11.13 &            13.90 \\
                                   Microbiology &  167964 & 2.04 &            4.11 &            10.97 &            14.47 \\
 Radiology, Nuclear Medicine \& Medical Imaging &  167725 & 2.04 &            2.31 &             5.25 &             6.65 \\
                    Endocrinology \& Metabolism &  165832 & 2.01 &            2.32 &             6.38 &             8.44 \\
   Public, Environmental \& Occupational Health &  151505 & 1.84 &            0.28 &             0.91 &             1.30 \\
                                     Physiology &  147080 & 1.79 &            1.16 &             3.54 &             4.97 \\
                                     Pediatrics &  137277 & 1.67 &            0.42 &             1.27 &             1.77 \\
                                     Psychiatry &  134227 & 1.63 &            0.70 &             1.69 &             2.36 \\
             Medicine, Research \& Experimental &  129280 & 1.57 &            4.60 &            11.36 &            14.26 \\
                                      Pathology &  119155 & 1.45 &            1.17 &             3.53 &             4.85 \\
                            Veterinary Sciences &  113394 & 1.38 &            0.70 &             2.19 &             3.13 \\
                                     Hematology &  111714 & 1.36 &            3.12 &             8.68 &            11.43 \\
                     Multidisciplinary Sciences &  110147 & 1.34 &           12.25 &            24.83 &            29.22 \\
                       Obstetrics \& Gynecology &  104895 & 1.27 &            0.95 &             2.60 &             3.64 \\
          Biotechnology \& Applied Microbiology &  101884 & 1.24 &            7.11 &            17.52 &            21.75 \\
                    Peripheral Vascular Disease &   99351 & 1.21 &            2.79 &             7.20 &             9.46 \\
                                     Toxicology &   98727 & 1.20 &            0.83 &             2.46 &             3.54 \\
                            Infectious Diseases &   98538 & 1.20 &            2.75 &             7.64 &            10.07 \\
                 Gastroenterology \& Hepatology &   95779 & 1.16 &            1.36 &             3.90 &             5.16 \\
\bottomrule
\end{tabular}
\end{table*}

These three steps leave us with $5\,461\,415$ unique papers, which is our final corpus, and the unit of our analysis is a paper. Table~\ref{tab:sc-cnt} provides the number and percentage of papers for the 30 most presented fields, which in total account for 70\% of all papers.

\subsection{Patent data}

To study technological impact of biomedical papers, we link them to the patented technology space and investigate if and to what extent they are cited as ``prior art''. Here we focus on patents granted at the U.S. Patent and Trademark Office (USPTO), with their bibliographic data sourced from \url{https://bulkdata.uspto.gov}. For a USPTO-granted patent, its front-page lists patent references as well as non-patent references (NPRs), both of which are considered as the ``prior art'' of the citing patent. While patent references cite previous patent literature, NPRs can refer to any type of documents including scientific papers. Our patent data cover all utility patents granted between 1976 and 2019, and we extract both types of references in these patents. Moreover, in our previous work \citep{Ke-compare-2018}, we have developed a highly-accurate (97\% accuracy) matching method to resolve a NPR to get whether and which MEDLINE paper it refers to, and the method was applied to all NPRs cited in USPTO patents granted between 1976 and 2012.

\subsection{Dependent variables}

Given the data of cited MEDLINE papers of all patents, we are able to collect the set of citing patents of a focal paper, from which we consider three categories of dependent variables that quantify the technological impact of the paper. The first category is binary variables indicating whether a paper has been cited by patents that are granted within the 5-, 10-, and 15-year windows after its publication. The main reason for looking at different lengths of citation-window is because the accumulation of patent citations is highly time-dependent. This can be readily seen from the summary statistics represented in Table~\ref{tab:var}, which indicates that while on average only 2.7\% of papers get cited by patents in 5 years, the percentage drastically increases to 9.1\% for 15-year citation window. Looking at individual fields, the rightmost three columns in Table~\ref{tab:sc-cnt} present the percentage of papers that obtain patent citations within 5, 10, and 15 years after publication. For Biochemistry \& Molecular Biology papers, only 5.5\% of them are cited by patents in 5 years, but the percentage more than tripled when we extend the citation window to 15 years, reaching to 18.4\%. The same trend is observed for many other fields, like Pharmacology \& Pharmacy, Immunology, and Cell Biology, etc.

\begin{table*}[t!]
\centering
\caption{Summary statistics of variables.\label{tab:var}}
\begin{tabular}{l c c c c r}
\toprule
Variable & Mean & Std. Dev. & Min & Max & N \\
\midrule
Cited by patents in 5 years & 0.027 & 0.163 & 0 & 1 & 5461415\\
Cited by patents in 10 years & 0.07 & 0.255 & 0 & 1 & 5461415\\
Cited by patents in 15 years & 0.091 & 0.287 & 0 & 1 & 5461415\\
5-year patent citations & 1.782 & 2.389 & 1 & 162 & 149643 \\
10-year patent citations & 2.989 & 6.079 & 1 & 552 & 381543 \\
15-year patent citations & 3.856 & 9.017 & 1 & 1572 & 495129 \\
5-year weighted patent citations & 14.332 & 52.494 & 0 & 5537 & 149643 \\
10-year weighted patent citations & 22.277 & 113.801 & 0 & 15719 & 381543 \\
15-year weighted patent citations & 28.231 & 148.370 & 0 & 19433 & 495129 \\
\midrule
Variety & 5.899 & 3.545 & 0 & 45 & 5461415\\
Balance & 0.699 & 0.271 & 0 & 1 & 5461415\\
Disparity & 0.378 & 0.185 & 0 & 0.999 & 5461415\\
Rao-Stirling & 0.235 & 0.139 & 0 & 0.77 & 5461415\\
\midrule
5-year scientific citations & 12.526 & 27.252 & 0 & 7240 & 5461415\\
10-year scientific citations & 22.015 & 53.488 & 0 & 30327 & 5461415\\
15-year scientific citations & 28.722 & 78.566 & 0 & 49315 & 5461415\\
Journal Impact Factor & 2.112 & 2.488 & 0 & 39.104 & 5461415\\
Number of MeSH terms & 11.655 & 4.832 & 1 & 57 & 5461415\\
Number of authors & 4.035 & 2.631 & 1 & 546 & 5461415\\
International collaboration & 0.106 & 0.308 & 0 & 1 & 4395138\\
Year & 1992.188 & 6.592 & 1980 & 2002 & 5461415\\
\bottomrule
\end{tabular}
\end{table*}

\begin{figure*}[t!]
\centering
\includegraphics[width=\textwidth]{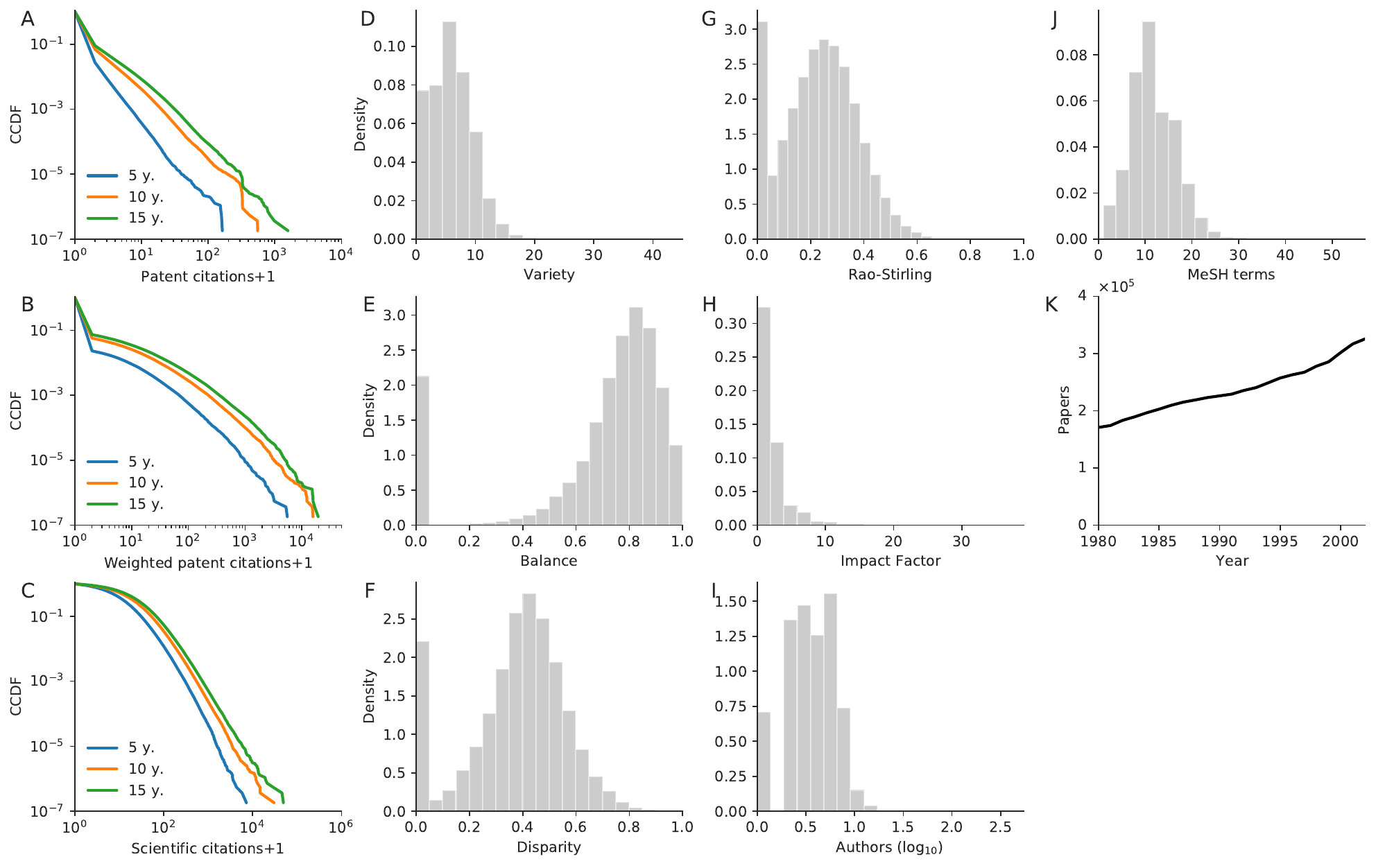}
\caption{Distribution of variables.}
\label{fig:var-dist}
\end{figure*}

For the subset of papers with patent citations, the second group of dependent variables is the number of citing patents in 5-, 10-, and 15-years. Figure~\ref{fig:var-dist}A plots the distributions of these variables, suggesting their over-dispersed feature and dependence on time.

Finally, considering that citing patents themselves have varying levels of technological impact upon followup technologies, the third group of dependent variables is the number of citing patents weighted by the number of forward citations those patents themselves obtained from other patents within 7 years after granting. Here we choose 7-year window because we are constrained by patents granted in 2012, which only have 7 years to accrue forward patent citations, since our patent-to-patent citation data cover patents until 2019. Figure~\ref{fig:var-dist}B plots the distributions of weighted patent citations, suggesting similarly that they are overly dispersed and the extent is intensified by technological impact of citing patents: The maximum weighted patent citations are more than ten times larger than their respective unweighted ones, as indicated from Table~\ref{tab:var}.

\subsection{Independent variables}

Our goal is to test the relationship between IDR and its impact upon patented technologies. Given the lack of consensus on a single, ``best'' indicator for interdisciplinarity \citep{Wagner-approache-2011, Wang-consistency-2020}, we employ several measures to quantify the extent of IDR of papers. These indicators build on the notion of integration of knowledge from different disciplines. Specifically, for each reference cited in the focal paper, we identify the discipline designations of the reference based on the journal where it was published. As a journal can belong to multiple SCs, we use the fractional counting procedure. That is, each reference takes the weight of one, which is equally split among the SCs to which the journal belongs. Aggregating over all references, we can calculate the fraction of each cited SC, denoted as $p_i$ for SC $i$.

Then, the first interdisciplinarity measure is variety, which is the number of SCs that the focal paper cites. Note that variety by our construction is not an integer number, as a cited field may be from only one reference in the focal paper, and the reference is associated with other SCs.

The second interdisciplinarity indicator is balance, which is the Shannon entropy diversity index of the probability distribution of cited SCs, normalized by the number of cited SCs, formally: 
\begin{equation}
\text{balance} = \frac{\text{entropy}} {\ln n} = - \frac{1}{\ln n} \sum_i p_i \ln p_i \,,
\end{equation}
where $n$ is the total number of unique SC cited. Note that for papers that cite only one SC, the balance is set to zero, since the entropy of cited SC is zero. Balance adds one layer of complexity to the variety measure, considering not only the number of fields cited but also the diversity in the distribution of those fields. A paper with a larger balance score indicates that it evenly cites references from across scientific fields. The measure has been widely used in existing literature \citep{Wang-impact-2015, Yegros-interdisc-2015}.

The third measure of interdisciplinarity is disparity, which is the average dissimilarity between two SCs:
\begin{equation}
\text{disparity} = \frac{2 \cdot \sum_{i < j} 1 - s_{ij}}{n(n-1)} \,.
\end{equation}
where $s_{ij}$ is the similarity between SC $i$ and $j$, and hence $1 - s_{ij}$ captures the dissimilarity between them. Following the literature, $s_{ij}$ is the cosine similarity between $i$ and $j$'s vectors of co-citations with other fields \citep{Porter-science-2009, Yegros-interdisc-2015}.

Finally, we also calculate the RS index, an integrated indicator of the three IDR measures, formally defined as:
\begin{equation}
\text{RS} = \sum_{i \neq j} p_i p_j (1 - s_{ij}) \,.
\end{equation}
The RS index adds further complexity, integrating variety, balance, and disparity between fields. Again, this indicator has been extensively used in  previous works \citep{Porter-science-2009, Cassi-analysing-2017, Gates-nature-2019}.

Figures~\ref{fig:var-dist}D--G show the distributions of the four IDR indicators.

\subsection{Control variables}

In light of previous literature, we consider several control variables. The first one is the Impact Factor (IF) of the journal where the focal paper was published, as publishing in high IF journals may increase visibility and readership, which may help expedite the knowledge flow to the technology domain. The second control variable is the number of Medical Subject Headings (MeSH) terms, which are controlled vocabularies used to indicate the topics of a paper. Different from other types of keywords that are added by authors, MeSH terms of a paper are chosen by trained staff at the National Library of Medicine. We include the number of MeSH terms to account for the consideration that topically more diverse papers may be more likely to be cited by patents due to wider applicability. MeSH terms and field designations of papers are two relatively independent characterizations of topics of papers: Papers belonging to the same field may have distinct MeSH terms, and some MeSH terms may be used to tag papers from disparate fields. Other control variables include the number of authors and whether the paper involves international collaboration. Figures~\ref{fig:var-dist}H--J plot the distributions of these control variables.

Furthermore, we consider the publication year and field fixed-effects and create dummy variables for each year and each SC. Thus the estimations capture within-year and within-field differences, meaning that the effects of IDR on technological impact are compared for papers in the same year and the same field. Year fixed-effect is included to control for some features, such as the number of citing patents, that are fixed in a year but change over time. Field fixed-effect is included, because there is an apparent field-dependent tendency of getting cited by patents for papers in different fields, as demonstrated in Table~\ref{tab:sc-cnt}. About 29\% of papers in the Multidisciplinary Sciences category have patent citations in 15 years. On the other extreme, less than 5\% of papers in several clinical medicine fields, such as General \& Internal Medicine and Surgery, get cited by patents. In between is Cell Biology, where 17\% of papers achieve technological impact. Note that for papers with multiple associated SCs, we pick the one that is most presented in the reference list, a procedure which has been used in the literature \citep{Ke-tech-2020, Verhoeven-measuring-2016}, and test the robustness of our results by using another procedure described in \ref{subsec:robust}.

For all the introduced variables, their summary statistics are reported in Table~\ref{tab:var}, their distributions are plotted in Figure~\ref{fig:var-dist}, and the correlations between them are presented in Table~\ref{tab:corr}.

\subsection{Regression methods}

In \ref{subsec:likelihood}, we use logistic regression to model the association between IDR and the likelihood of technological impact, and the model specification is as follows:
\begin{equation} \label{eq:reg}
c_i = \beta_0 + \beta_1 \cdot IDR_i + \gamma \cdot \text{controls}_i + \delta_i + \eta_i + \varepsilon_i \, ,
\end{equation}
where $c_i$ indicates whether paper $i$ has been cited by patents, $\beta_0$ is the intercept, $\beta_1$ is the coefficients of interest for the IDR independent variables, $\gamma$ represents coefficients for control variables, $\delta_i$ is a dummy variable for WoS SC (field fixed effect), $\eta_i$ is a dummy variable for publication year (year fixed effect), and $\varepsilon_i$ is the noise term.

In \ref{subsec:intensity} and \ref{subsec:weighted}, we employ negative binomial regression for modeling the effects of IDR on unweighted and weighted intensity of technological impact ($c_i$ in Eq.~\ref{eq:reg}), as both are over-dispersed variables (Table~\ref{tab:var}). The model specifications are similar to the logistic regression one.

\section{Results} \label{sec:res}

\subsection{Likelihood of technological impact} \label{subsec:likelihood}

We employ logistic regression to model the effect of the extent of IDR of a paper on its likelihood of getting cited by patents. Table~\ref{tab:citedbypat10} presents the modeling results for the case of 10-year citation window, where the dependent variable is whether a paper has been cited by patents that are granted within 10 years after the publication of the paper. Model~1 is the baseline model where we only consider control variables. Models 2--10 include our independent variables of IDR indicators. The reductions of the BIC of these models from the BIC of Model 1 lend very strong support for these models and hence the explanatory power of the IDR measures.

Model~2 includes the RS index and indicates its positive, statistically significant relationship with the likelihood of having technological impact. After controlling for confounders, a one standard deviation increase in RS translates to, on average, a 0.146 increase in the log odds of getting patent citations. 

\begin{landscape}
\begin{table*}
\centering
\caption{Logistic regression modeling of whether a paper has patent citations in 10 years. \label{tab:citedbypat10}}
\begin{tabular}{l*{10}{c}}
\toprule%
& (1) & (2) & (3) & (4) & (5) & (6) & (7) & (8) & (9) & (10) \\
\midrule
JIF                 &       0.149\sym{***}&       0.155\sym{***}&       0.145\sym{***}&       0.150\sym{***}&       0.151\sym{***}&       0.142\sym{***}&       0.156\sym{***}&       0.139\sym{***}&       0.139\sym{***}&       0.142\sym{***}\\
                    &  (0.000653)         &  (0.000663)         &  (0.000652)         &  (0.000655)         &  (0.000658)         &  (0.000657)         &  (0.000665)         &  (0.000658)         &  (0.000661)         &  (0.000657)         \\
MeSH                &      0.0419\sym{***}&      0.0434\sym{***}&      0.0320\sym{***}&      0.0413\sym{***}&      0.0427\sym{***}&      0.0306\sym{***}&      0.0426\sym{***}&      0.0284\sym{***}&      0.0288\sym{***}&      0.0307\sym{***}\\
                    &  (0.000387)         &  (0.000388)         &  (0.000395)         &  (0.000388)         &  (0.000388)         &  (0.000400)         &  (0.000390)         &  (0.000402)         &  (0.000403)         &  (0.000401)         \\
Authors ($\ln$)     &       0.467\sym{***}&       0.444\sym{***}&       0.421\sym{***}&       0.448\sym{***}&       0.456\sym{***}&       0.420\sym{***}&       0.437\sym{***}&       0.415\sym{***}&       0.419\sym{***}&       0.421\sym{***}\\
                    &   (0.00353)         &   (0.00355)         &   (0.00356)         &   (0.00355)         &   (0.00354)         &   (0.00357)         &   (0.00356)         &   (0.00358)         &   (0.00358)         &   (0.00358)         \\
RS                  &                     &       1.052\sym{***}&                     &                     &                     &                     &       2.229\sym{***}&                     &                     &                     \\
                    &                     &    (0.0156)         &                     &                     &                     &                     &    (0.0499)         &                     &                     &                     \\
Variety             &                     &                     &      0.0856\sym{***}&                     &                     &      0.0875\sym{***}&                     &       0.199\sym{***}&      0.0871\sym{***}&      0.0877\sym{***}\\
                    &                     &                     &  (0.000587)         &                     &                     &  (0.000661)         &                     &   (0.00231)         &  (0.000664)         &  (0.000668)         \\
Balance             &                     &                     &                     &       0.738\sym{***}&                     &       0.416\sym{***}&                     &       0.215\sym{***}&       1.796\sym{***}&       0.426\sym{***}\\
                    &                     &                     &                     &    (0.0106)         &                     &    (0.0118)         &                     &    (0.0127)         &    (0.0428)         &    (0.0128)         \\
Disparity           &                     &                     &                     &                     &       0.474\sym{***}&      -0.506\sym{***}&                     &      -0.636\sym{***}&      -0.692\sym{***}&      -0.605\sym{***}\\
                    &                     &                     &                     &                     &    (0.0125)         &    (0.0151)         &                     &    (0.0155)         &    (0.0161)         &    (0.0511)         \\
$\text{RS}^2$       &                     &                     &                     &                     &                     &                     &      -2.303\sym{***}&                     &                     &                     \\
                    &                     &                     &                     &                     &                     &                     &    (0.0927)         &                     &                     &                     \\
$\text{Variety}^2$  &                     &                     &                     &                     &                     &                     &                     &    -0.00638\sym{***}&                     &                     \\
                    &                     &                     &                     &                     &                     &                     &                     &  (0.000127)         &                     &                     \\
$\text{Balance}^2$  &                     &                     &                     &                     &                     &                     &                     &                     &      -1.200\sym{***}&                     \\
                    &                     &                     &                     &                     &                     &                     &                     &                     &    (0.0354)         &                     \\
$\text{Disparity}^2$&                     &                     &                     &                     &                     &                     &                     &                     &                     &       0.131\sym{*}  \\
                    &                     &                     &                     &                     &                     &                     &                     &                     &                     &    (0.0642)         \\
Constant            &      -3.316\sym{***}&      -3.700\sym{***}&      -3.963\sym{***}&      -3.884\sym{***}&      -3.532\sym{***}&      -4.065\sym{***}&      -3.753\sym{***}&      -4.104\sym{***}&      -4.223\sym{***}&      -4.062\sym{***}\\
                    &     (0.132)         &     (0.132)         &     (0.134)         &     (0.132)         &     (0.132)         &     (0.134)         &     (0.132)         &     (0.133)         &     (0.134)         &     (0.134)         \\
\hline
Field fe            & $\checkmark$        & $\checkmark$        & $\checkmark$        & $\checkmark$        & $\checkmark$        & $\checkmark$        & $\checkmark$        & $\checkmark$        & $\checkmark$        & $\checkmark$ \\
Year fe             & $\checkmark$        & $\checkmark$        & $\checkmark$        & $\checkmark$        & $\checkmark$        & $\checkmark$        & $\checkmark$        & $\checkmark$        & $\checkmark$        & $\checkmark$ \\
Observations        &     5453416         &     5453416         &     5453416         &     5453416         &     5453416         &     5453416         &     5453416         &     5453416         &     5453416         &     5453416         \\
Pseudo \(R^{2}\)    &       0.158         &       0.160         &       0.166         &       0.160         &       0.159         &       0.166         &       0.160         &       0.167         &       0.167         &       0.166         \\
\textit{BIC}        &   2330529.6         &   2325991.4         &   2309755.8         &   2325265.7         &   2329079.0         &   2307957.1         &   2325377.2         &   2305288.6         &   2306800.5         &   2307968.5         \\
\bottomrule
\multicolumn{5}{l}{\footnotesize Standard errors in parentheses}\\
\multicolumn{5}{l}{\footnotesize \sym{*} \(p<0.05\), \sym{**} \(p<0.01\), \sym{***} \(p<0.001\)}\\
\end{tabular}
\end{table*}
\end{landscape}

Models 3--5 focus on each of the three dimensions of IDR separately. We find that all the three dimensions have positive, statistically significant associations with the likelihood of being cited by patents. Model~3 suggests that after controlling for confounders, citing one more field is associated with a 8.9\% increase in the odds of getting cited by patents. Model~4 shows that the effect size of the positive association between balance and likelihood of technological impact is pronounced; a one standard deviation increase in balance produces, on average, a 0.2 increase in the log odds of getting patent citations. Model~5 shows that for a one standard deviation increase in disparity, we expect a 0.088 increase in the log odds of receiving patent citations, holding all control variables constant.

Model~6 examines the three aspects of IDR together, indicating that after controlling for each other, the directions of the associations with technological impact still persist for variety and balance, whereas disparity now negatively affects technological impact. The effect size of variety remains similar to the Model~3 case---9.1\% increase in the odds for citing one more field. The effect size of balance halves from Model~4; a one standard deviation increase in balance is linked to a 0.11 increase in the log odds of getting cited by patents. A one standard deviation increase of disparity is associated with a 0.093 decrease in the log odds. The reason why the association for disparity is negative may be because there is a relatively strong correlation between disparity and variety (coefficient 0.454) and balance (coefficient 0.628).

\begin{figure*}[t!]
\centering
\includegraphics[width=0.495\textwidth]{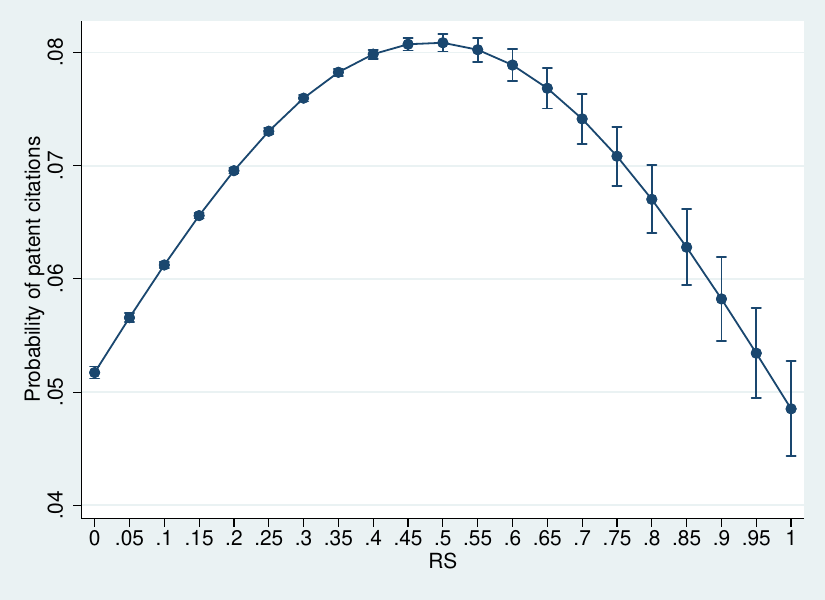}
\includegraphics[width=0.495\textwidth]{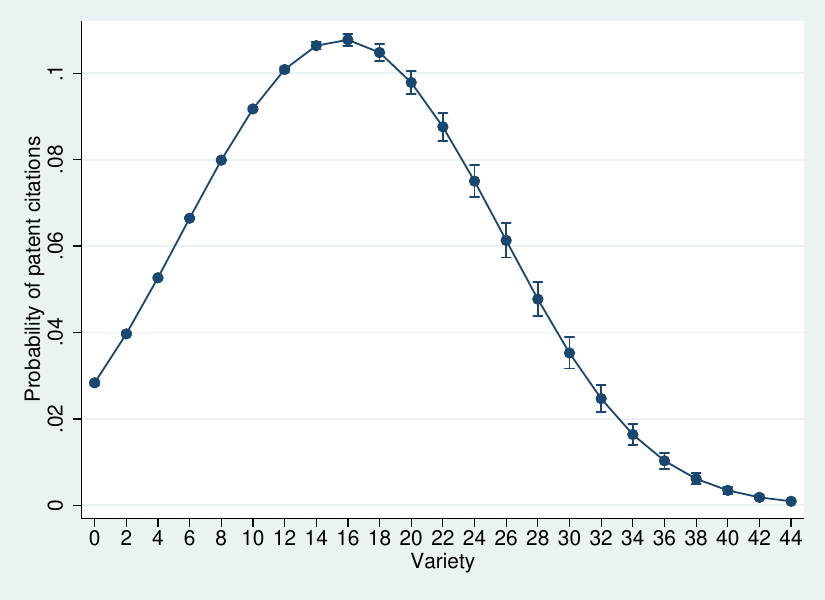}
\includegraphics[width=0.495\textwidth]{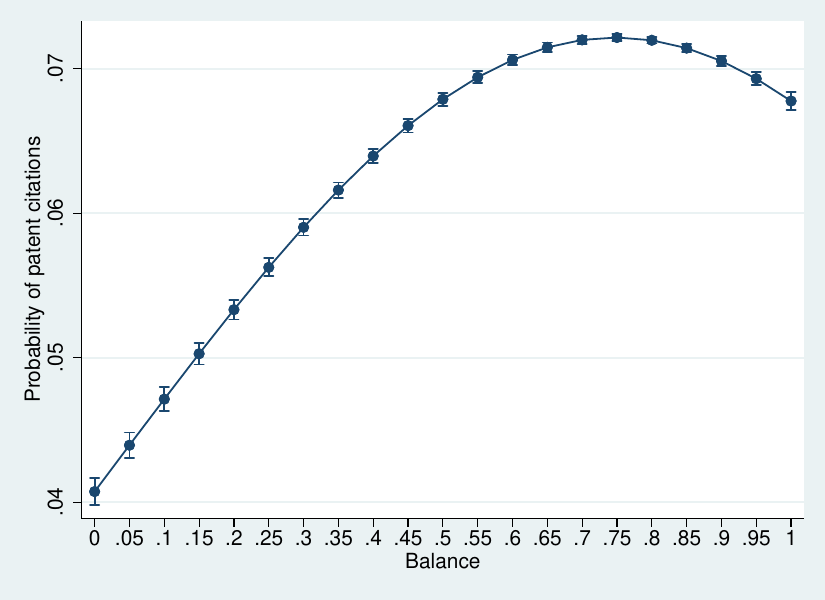}
\includegraphics[width=0.495\textwidth]{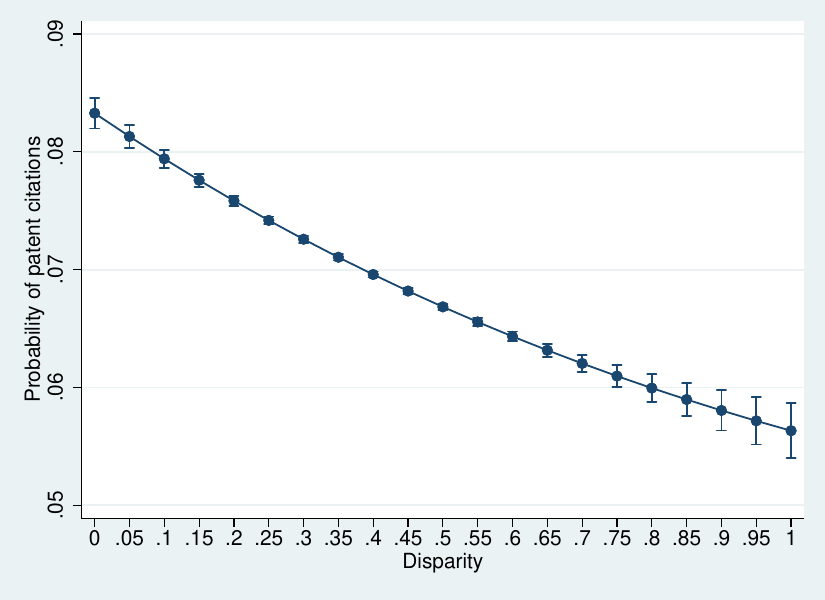}
\caption{Estimated probabilities of getting cited by patents in 10-years.}
\label{fig:margin}
\end{figure*}

While Models 3--6 capture linear relationships between IDR and likelihood of technological impact, curvilinear ones may also potentially be present, considering that recent studies have emphasized curvilinear relationships between IDR, particularly the disparity aspect, and scientific impact \citep{Leahey-prominent-2017, Yegros-interdisc-2015}. Therefore, we explore curvilinear links between IDR and technological impact, by adding a quadratic term for each of the four IDR measures to Model~6. Model~7 shows that the quadratic term for RS is statistically significant and negative, implying the presence of a curvilinear relationship between RS and the likelihood of achieving technological impact. Such a relationship is illustrated in Figure~\ref{fig:margin}A, which plots predictive margins as RS increases, demonstrating a positive effect of RS on the probability of getting cited by patents can only be realized up to a threshold, beyond which RS negatively affects the chance of obtaining patent citations. However, a closer inspection of Figure~\ref{fig:margin}A reveals that the negative side of the curvilinear relationship occurs when RS gets large (RS $\geq 0.5$), corresponding to the region where only a rather small fraction of papers fall into, as evidenced from Figure~\ref{fig:var-dist}G, which presents the distribution of RS. This means that for most papers, increasing RS would increase their probability of attaining technological impact, in agreement with the positive coefficient for RS in Models~2.

Similarly, Model~8 shows a negative, statistically significant coefficient for the quadratic term of variety, suggesting the presence of a curvilinear relationship between variety and likelihood of patent citations, as shown in Figure~\ref{fig:margin}B. Again, the downward trend part of the curve starts from when variety is larger than around 16, a region where there are only a small fraction of papers (\emph{cf}. Figure~\ref{fig:var-dist}D), indicating that a simpler linear relationship may be enough to capture the association between variety and likelihood of technological impact.

The case for balance is different from the previous two IDR measures. Model~9 reports a statistically significant, negative coefficient for the quadratic term for balance, signifying the existence of a curvilinear relationship between balance and the likelihood of technological impact, as illustrated in Figure~\ref{fig:margin}C. It presents that the probability of patent citations keeps increasing until when balance reaches around 0.75, based on a visual inspection. This upward trend part accounts for 41\% papers (\emph{cf}. Figure~\ref{fig:var-dist}E), indicating that the turning point is located well within the range of balance (between 0 and 1) and pointing to the potential presence of inverted U-shaped relationship. To test this formally, we employ the three-step testing procedure proposed in \citet{lind2010or}, following \citet{haans2016thinking}. The first step is a significant and negative coefficient for the quadratic term, which is satisfied. The second step is that the slopes, which is given by $\beta_1 + 2\beta_2X$ ($\beta_2$ is the coefficient for the quadratic term.), at both end-points of the data range are sufficiently steep (significantly different from zero). That is, in the case of inverted U-shape, the slope at the low end should be significantly positive and at the high end significantly negative. In our case, the slope at the low end (balance at 0) is $1.796$, and the slope at the high end (balance at 1) is $-0.604$. Both are significant, with their $p$-values close to zero. Thus, the second condition is met. The third step is that the turning point, which is $-\beta_1/(2\beta_2)$, should be well located within the data range. Here, the turning point for balance is 0.748, and the 95\% confidence is $[0.734, 0.763]$ \citep{fieller1954some}, which is within the data range. Based on these results, we conclude that there exists an inverted U-shaped relationship between balance and the likelihood of technological impact.

For the case of disparity, although Model~10 presents a statistically significant, positive coefficient for the quadratic term for disparity, a curvilinear relationship between disparity and the likelihood of technological impact is not clearly present: Figure~\ref{fig:margin}D displays that the probability of getting cited by patents keeps decreasing, without an obvious uptrend part, consistent with the negative coefficient for disparity in Model~6.

Table~\ref{tab:citedbypat10} also reveals that (1) papers published in high IF journals are more likely to get patent citations; (2) papers that are topically more diverse, as measured by number of MeSH terms, are more superior in generating technological impact; and (3) the number of authors is positively linked to the likelihood of patent citations, consistent with its positive effect on scientific impact.

To summarize, we find that the RS index has a positive association with likelihood of technological impact. For the three dimensions of IDR, the number of fields a paper cites and the evenness of the distribution over those cited fields have positive effects on the likelihood of being cited by patents, but both effects can be offset if the paper draws upon distant fields. 

\subsection{Intensity of technological impact} \label{subsec:intensity}

We have looked at whether papers are cited by patents. We now focus on the number of patent citations and examine if it is affected by the three aspects of IDR or the RS index. We restrict this analysis to the subset of papers in our corpus that have gained technological impact. We use negative binomial regression, since the number of patent citations is an over-dispersed variable (Table~\ref{tab:var}) and heterogeneously distributed (Figure~\ref{fig:var-dist}), with one paper getting cited by 552 patents in 10 years.

\begin{landscape}
\begin{table*}[t!]
\centering
\caption{Negative binomial regression modeling of 10-year patent citations. \label{tab:patc10}}
\begin{tabular}{l*{10}{c}}
\toprule%
& (1) & (2) & (3) & (4) & (5) & (6) & (7) & (8) & (9) & (10) \\
\midrule
JIF                 &      0.0344\sym{***}&      0.0349\sym{***}&      0.0344\sym{***}&      0.0345\sym{***}&      0.0347\sym{***}&      0.0347\sym{***}&      0.0349\sym{***}&      0.0347\sym{***}&      0.0349\sym{***}&      0.0346\sym{***}\\
                    &  (0.000436)         &  (0.000440)         &  (0.000436)         &  (0.000437)         &  (0.000439)         &  (0.000439)         &  (0.000441)         &  (0.000440)         &  (0.000442)         &  (0.000440)         \\
MeSH                &   -0.000255         &   0.0000679         &   -0.000554         &   -0.000245         &  -0.0000343         &   -0.000216         &    0.000100         &   -0.000187         &  -0.0000280         &  -0.0000959         \\
                    &  (0.000334)         &  (0.000335)         &  (0.000337)         &  (0.000334)         &  (0.000335)         &  (0.000342)         &  (0.000336)         &  (0.000344)         &  (0.000344)         &  (0.000343)         \\
Authors ($\ln$)     &       0.125\sym{***}&       0.122\sym{***}&       0.123\sym{***}&       0.123\sym{***}&       0.124\sym{***}&       0.122\sym{***}&       0.123\sym{***}&       0.122\sym{***}&       0.122\sym{***}&       0.123\sym{***}\\
                    &   (0.00298)         &   (0.00299)         &   (0.00299)         &   (0.00298)         &   (0.00299)         &   (0.00300)         &   (0.00300)         &   (0.00300)         &   (0.00300)         &   (0.00300)         \\
RS                  &                     &       0.137\sym{***}&                     &                     &                     &                     &      0.0642         &                     &                     &                     \\
                    &                     &    (0.0147)         &                     &                     &                     &                     &    (0.0458)         &                     &                     &                     \\
Variety             &                     &                     &     0.00322\sym{***}&                     &                     &    0.000768         &                     &   -0.000969         &    0.000833         &     0.00130\sym{*}  \\
                    &                     &                     &  (0.000544)         &                     &                     &  (0.000606)         &                     &   (0.00206)         &  (0.000606)         &  (0.000616)         \\
Balance             &                     &                     &                     &       0.135\sym{***}&                     &       0.120\sym{***}&                     &       0.122\sym{***}&     -0.0661         &       0.135\sym{***}\\
                    &                     &                     &                     &    (0.0109)         &                     &    (0.0116)         &                     &    (0.0121)         &    (0.0410)         &    (0.0121)         \\
Disparity           &                     &                     &                     &                     &      0.0860\sym{***}&      0.0395\sym{**} &                     &      0.0413\sym{**} &      0.0612\sym{***}&      -0.170\sym{***}\\
                    &                     &                     &                     &                     &    (0.0121)         &    (0.0135)         &                     &    (0.0137)         &    (0.0143)         &    (0.0454)         \\
$\text{RS}^2$       &                     &                     &                     &                     &                     &                     &       0.139         &                     &                     &                     \\
                    &                     &                     &                     &                     &                     &                     &    (0.0829)         &                     &                     &                     \\
$\text{Variety}^2$  &                     &                     &                     &                     &                     &                     &                     &   0.0000974         &                     &                     \\
                    &                     &                     &                     &                     &                     &                     &                     &  (0.000111)         &                     &                     \\
$\text{Balance}^2$  &                     &                     &                     &                     &                     &                     &                     &                     &       0.157\sym{***}&                     \\
                    &                     &                     &                     &                     &                     &                     &                     &                     &    (0.0333)         &                     \\
$\text{Disparity}^2$&                     &                     &                     &                     &                     &                     &                     &                     &                     &       0.277\sym{***}\\
                    &                     &                     &                     &                     &                     &                     &                     &                     &                     &    (0.0573)         \\
Constant            &       0.305         &       0.251         &       0.288         &       0.193         &       0.256         &       0.180         &       0.257         &       0.183         &       0.206         &       0.189         \\
                    &     (0.738)         &     (0.739)         &     (0.738)         &     (0.739)         &     (0.738)         &     (0.739)         &     (0.739)         &     (0.739)         &     (0.739)         &     (0.739)         \\
\hline
lnalpha             &      -0.544\sym{***}&      -0.544\sym{***}&      -0.544\sym{***}&      -0.544\sym{***}&      -0.544\sym{***}&      -0.544\sym{***}&      -0.544\sym{***}&      -0.544\sym{***}&      -0.545\sym{***}&      -0.545\sym{***}\\
                    &   (0.00317)         &   (0.00317)         &   (0.00317)         &   (0.00317)         &   (0.00317)         &   (0.00317)         &   (0.00317)         &   (0.00317)         &   (0.00317)         &   (0.00317)         \\
\hline
Field fe            & $\checkmark$        & $\checkmark$        & $\checkmark$        & $\checkmark$        & $\checkmark$        & $\checkmark$        & $\checkmark$        & $\checkmark$        & $\checkmark$        & $\checkmark$        \\
Year fe             & $\checkmark$        & $\checkmark$        & $\checkmark$        & $\checkmark$        & $\checkmark$        & $\checkmark$        & $\checkmark$        & $\checkmark$        & $\checkmark$        & $\checkmark$        \\
\(N\)               &      381543         &      381543         &      381543         &      381543         &      381543         &      381543         &      381543         &      381543         &      381543         &      381543         \\
\textit{BIC}        &   1654997.7         &   1654923.3         &   1654975.5         &   1654857.7         &   1654960.1         &   1654868.9         &   1654933.3         &   1654881.0         &   1654859.4         &   1654858.3         \\
\bottomrule
\multicolumn{5}{l}{\footnotesize Standard errors in parentheses}\\
\multicolumn{5}{l}{\footnotesize \sym{*} \(p<0.05\), \sym{**} \(p<0.01\), \sym{***} \(p<0.001\)}\\
\end{tabular}
\end{table*}
\end{landscape}

Table~\ref{tab:patc10} presents the modeling results for 10-year patent citations, which indicate positive and statistically significant associations between IDR and number of patent citations. Model~2 implies that if a paper's RS index were to increase by a one standard deviation, we would expect its patent citations to increase by 2\%. This linkage is consistent with the linkage to likelihood of technological impact. Taken together, Tables~\ref{tab:citedbypat10} and \ref{tab:patc10} suggest that the probability of a paper obtaining patent citations increases with its RS index, and conditional on getting cited by patents, the number of patent citations also increases with RS index.

Models~3--6 assess the linear relationships between the three IDR indicators and intensity of technological impact. Model~3 shows that variety has a positive, but very small, effect on the number of patent citations, to the extent that after controlling for the other two aspects of IDR, there is a lack of significance, as seen from Model~6. Citing one more field means that 10-year patent citations increase by less than half percent (Model~1). Models~4 and~6 suggest that balance is positively linked to number of patent citations, even after controlling for variety and disparity, with one standard deviation increase in balance translating to a significant 3.2\% increase in the number of 10-year patent citations (Model~6). This correlation is in the same direction as the correlation with likelihood of patent citations. Models~5 and~6 indicate that disparity also has a positive linkage to patent citations, though the effect size is much smaller than that of balance. A 0.73\% increase in the number of patent citations is expected if we were to increase the disparity of a paper by one standard deviation (Model~6).

Models~7--10 test the potential presence of curvilinear relationships between IDR and number of patent citations. The lack of statistical significance for the quadratic terms for the RS index and variety indicates the lack of curvilinear relationships, whereas the positive and statistically significant quadratic terms for balance and disparity suggest that both correlations with intensity of technological impact exhibit curvilinear relationships.

\subsection{Weighted intensity of technological impact} \label{subsec:weighted}

\begin{landscape}
\begin{table*}[t!]
\centering
\caption{Negative binomial regression modeling of weighted 10-year patent citations. \label{tab:wpatc10}}
\begin{tabular}{l*{10}{c}}
\toprule%
& (1) & (2) & (3) & (4) & (5) & (6) & (7) & (8) & (9) & (10) \\
\midrule
JIF                 &      0.0303\sym{***}&      0.0322\sym{***}&      0.0306\sym{***}&      0.0304\sym{***}&      0.0325\sym{***}&      0.0341\sym{***}&      0.0312\sym{***}&      0.0352\sym{***}&      0.0355\sym{***}&      0.0338\sym{***}\\
                    &  (0.000802)         &  (0.000811)         &  (0.000804)         &  (0.000802)         &  (0.000809)         &  (0.000816)         &  (0.000811)         &  (0.000822)         &  (0.000826)         &  (0.000817)         \\
MeSH                &     -0.0237\sym{***}&     -0.0223\sym{***}&     -0.0222\sym{***}&     -0.0237\sym{***}&     -0.0221\sym{***}&     -0.0183\sym{***}&     -0.0213\sym{***}&     -0.0169\sym{***}&     -0.0169\sym{***}&     -0.0166\sym{***}\\
                    &  (0.000576)         &  (0.000581)         &  (0.000584)         &  (0.000576)         &  (0.000579)         &  (0.000594)         &  (0.000583)         &  (0.000598)         &  (0.000599)         &  (0.000597)         \\
Authors ($\ln$)     &      0.0636\sym{***}&      0.0528\sym{***}&      0.0732\sym{***}&      0.0615\sym{***}&      0.0526\sym{***}&      0.0659\sym{***}&      0.0629\sym{***}&      0.0707\sym{***}&      0.0663\sym{***}&      0.0761\sym{***}\\
                    &   (0.00514)         &   (0.00516)         &   (0.00517)         &   (0.00515)         &   (0.00515)         &   (0.00517)         &   (0.00517)         &   (0.00518)         &   (0.00517)         &   (0.00518)         \\
RS                  &                     &       0.493\sym{***}&                     &                     &                     &                     &      -1.364\sym{***}&                     &                     &                     \\
                    &                     &    (0.0244)         &                     &                     &                     &                     &    (0.0791)         &                     &                     &                     \\
Variety             &                     &                     &     -0.0157\sym{***}&                     &                     &     -0.0314\sym{***}&                     &      -0.103\sym{***}&     -0.0314\sym{***}&     -0.0271\sym{***}\\
                    &                     &                     &  (0.000929)         &                     &                     &   (0.00103)         &                     &   (0.00363)         &   (0.00103)         &   (0.00104)         \\
Balance             &                     &                     &                     &       0.105\sym{***}&                     &      0.0326         &                     &       0.162\sym{***}&      -1.103\sym{***}&       0.227\sym{***}\\
                    &                     &                     &                     &    (0.0182)         &                     &    (0.0192)         &                     &    (0.0205)         &    (0.0685)         &    (0.0207)         \\
Disparity           &                     &                     &                     &                     &       0.540\sym{***}&       0.807\sym{***}&                     &       0.888\sym{***}&       0.951\sym{***}&      -1.402\sym{***}\\
                    &                     &                     &                     &                     &    (0.0197)         &    (0.0227)         &                     &    (0.0232)         &    (0.0242)         &    (0.0788)         \\
$\text{RS}^2$       &                     &                     &                     &                     &                     &                     &       3.510\sym{***}&                     &                     &                     \\
                    &                     &                     &                     &                     &                     &                     &     (0.142)         &                     &                     &                     \\
$\text{Variety}^2$  &                     &                     &                     &                     &                     &                     &                     &     0.00393\sym{***}&                     &                     \\
                    &                     &                     &                     &                     &                     &                     &                     &  (0.000194)         &                     &                     \\
$\text{Balance}^2$  &                     &                     &                     &                     &                     &                     &                     &                     &       0.984\sym{***}&                     \\
                    &                     &                     &                     &                     &                     &                     &                     &                     &    (0.0564)         &                     \\
$\text{Disparity}^2$&                     &                     &                     &                     &                     &                     &                     &                     &                     &       2.920\sym{***}\\
                    &                     &                     &                     &                     &                     &                     &                     &                     &                     &    (0.0998)         \\
Constant            &       1.848         &       1.649         &       1.928         &       1.760         &       1.542         &       1.524         &       1.809         &       1.647         &       1.664         &       1.608         \\
                    &     (1.201)         &     (1.201)         &     (1.201)         &     (1.201)         &     (1.200)         &     (1.199)         &     (1.200)         &     (1.198)         &     (1.199)         &     (1.198)         \\
\hline
lnalpha             &       1.021\sym{***}&       1.020\sym{***}&       1.020\sym{***}&       1.020\sym{***}&       1.019\sym{***}&       1.017\sym{***}&       1.018\sym{***}&       1.016\sym{***}&       1.016\sym{***}&       1.015\sym{***}\\
                    &   (0.00219)         &   (0.00219)         &   (0.00219)         &   (0.00219)         &   (0.00219)         &   (0.00219)         &   (0.00219)         &   (0.00219)         &   (0.00219)         &   (0.00219)         \\
\hline
Field fe            & $\checkmark$        & $\checkmark$        & $\checkmark$        & $\checkmark$        & $\checkmark$        & $\checkmark$        & $\checkmark$        & $\checkmark$        & $\checkmark$        & $\checkmark$        \\
Year fe             & $\checkmark$        & $\checkmark$        & $\checkmark$        & $\checkmark$        & $\checkmark$        & $\checkmark$        & $\checkmark$        & $\checkmark$        & $\checkmark$        & $\checkmark$        \\
\(N\)               &      381543         &      381543         &      381543         &      381543         &      381543         &      381543         &      381543         &      381543         &      381543         &      381543         \\
\textit{BIC}        &   2773890.8         &   2773494.7         &   2773619.5         &   2773870.4         &   2773160.0         &   2772256.2         &   2772876.1         &   2771826.6         &   2771959.0         &   2771379.8         \\
\bottomrule
\multicolumn{5}{l}{\footnotesize Standard errors in parentheses}\\
\multicolumn{5}{l}{\footnotesize \sym{*} \(p<0.05\), \sym{**} \(p<0.01\), \sym{***} \(p<0.001\)}\\
\end{tabular}
\end{table*}
\end{landscape}

\begin{figure*}[t!]
\centering
\includegraphics[width=0.495\textwidth]{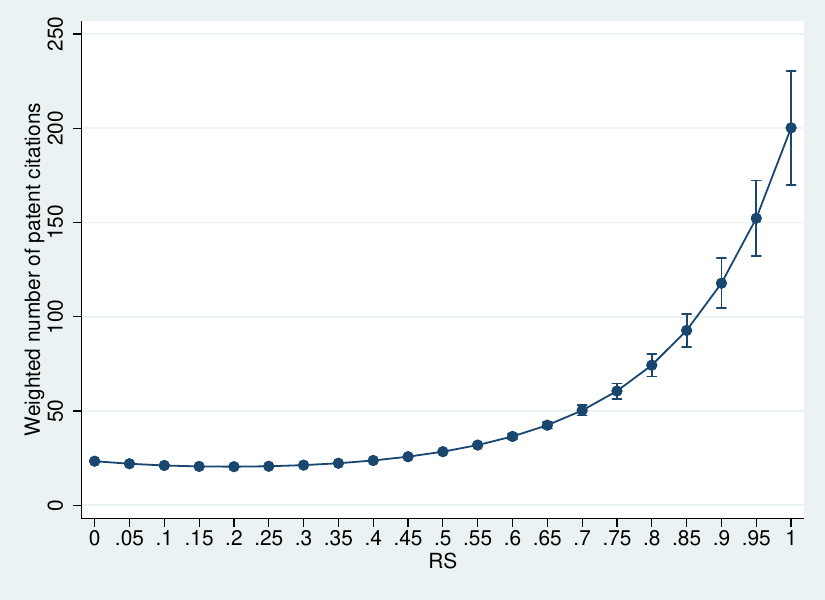}
\includegraphics[width=0.495\textwidth]{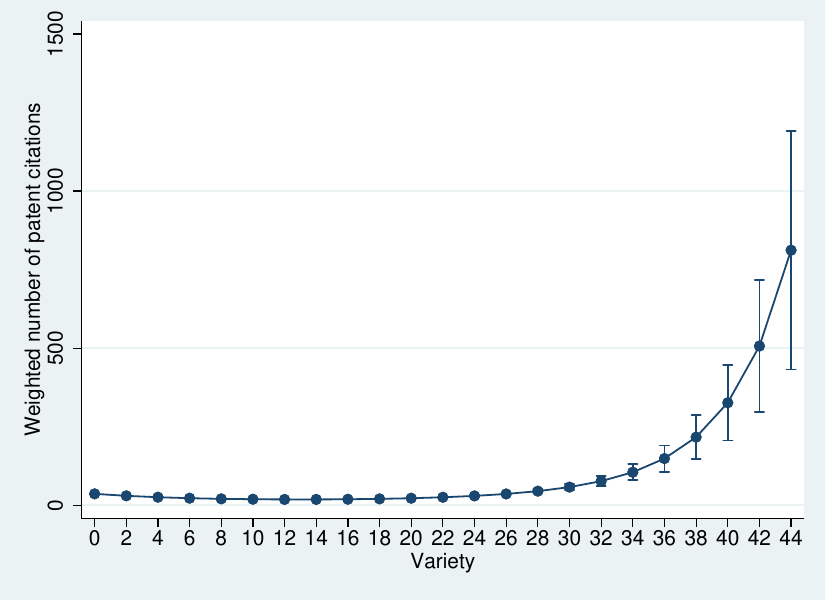}
\includegraphics[width=0.495\textwidth]{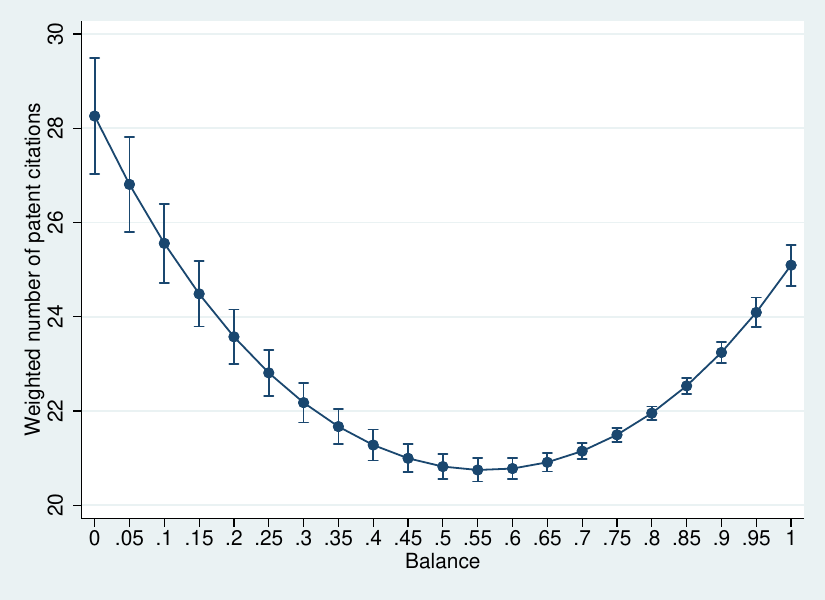}
\includegraphics[width=0.495\textwidth]{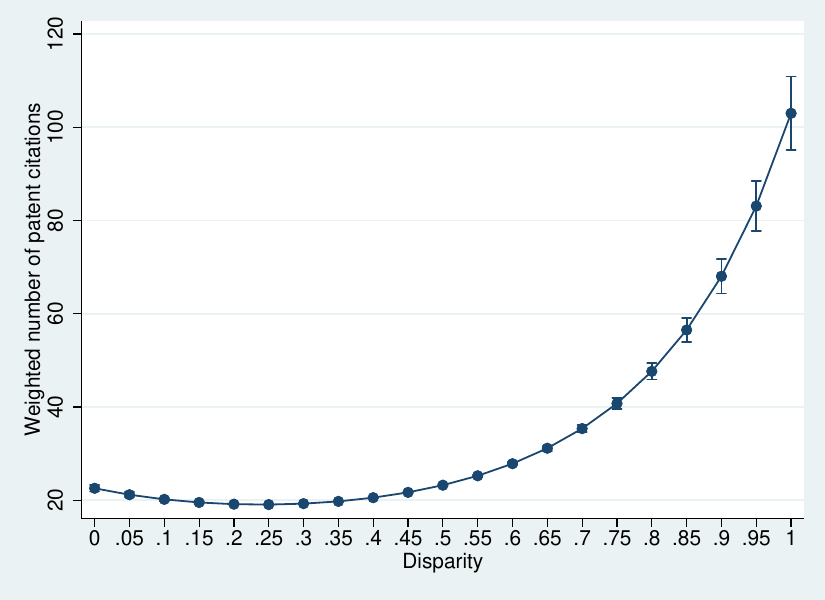}
\caption{Estimated weighted number of patent citations in 10-years.}
\label{fig:margin:wpat10}
\end{figure*}

Finally, we examine weighted intensity of technological impact of papers, considering not only the number of citing patents but also technological impact of those patents themselves. Table~\ref{tab:wpatc10} presents the modeling results from negative binomial regression for 10-year citation window. Looking at the RS index, Model~2 demonstrates that it has a positive, sizeable effect on weighted number of patent citations; we expect an increase by 6.8\% for a one standard deviation increase in RS, which is more than three times larger than the size presented in the unweighted case. Model~7 further includes the quadratic term, showing the presence of a curvilinear relationship, as presented in Figure~\ref{fig:margin:wpat10}A. For variety, Models~3 and~6 show that it bears a negative effect; after controlling for confounders, citing one more field is associated with a decrease of the weighted number of patent citations by 3\%. Model~8 reports a positive and statistically significant coefficient for the quadratic term of variety, suggesting a curvilinear relationship that is illustrated in Figure~\ref{fig:margin:wpat10}B. Turning to balance, it carries a small, statistically insignificant effect, after controlling for the other two dimensions of IDR (Model~6). Model~9 and Figure~\ref{fig:margin:wpat10}C indicate the presence of a curvilinear relationship between balance and number of weighted patent citations. Finally, disparity is largely correlated with weighted number of patent citations; a one standard deviation increase in disparity is linked to a 14.9\% increase in weighted citations (Model~6). Though Model~10 reports a curvilinear linkage between disparity and weighted patent citations, a visual inspection into Figure~\ref{fig:margin:wpat10}D reveals the lack of U-shaped relationship.

\subsection{Robustness tests} \label{subsec:robust}

\begin{figure*}[t!]
\centering
\includegraphics[width=\textwidth]{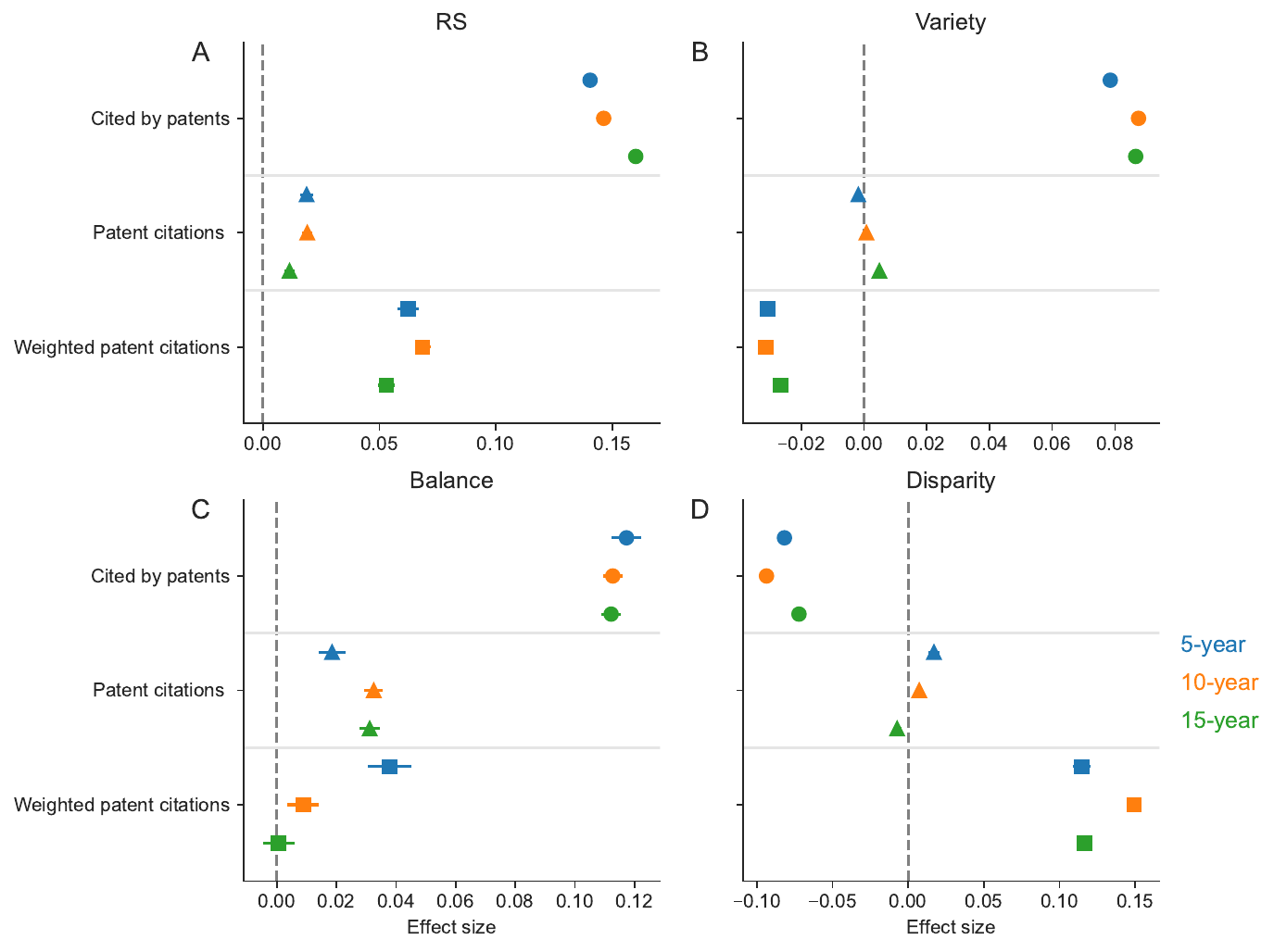}
\caption{The effect sizes of RS, variety, balance, and disparity.}
\label{fig:coeff}
\end{figure*}

We perform several additional tests to examine the robustness of our results. First, our results thus far are based on a citation-window of 10 years after the publication of a paper. We further analyze if and to what extent the effects of IDR on technological impact are dependent on the length of citation window. To this end, we repeat our analyses for 5- and 15-year windows, and the results are reported in Tables~\ref{tab:citedbypat5}--\ref{tab:wpatc15}. To facilitate comparisons of the effects based on different window lengths, Figure~\ref{fig:coeff} plots the effect sizes of the four IDR metrics, suggesting that window length does not have a significant influence on the effects of IDR. Figure~\ref{fig:coeff}A focuses on the RS index, where effect size is measured in terms of one standard deviation increase in RS in Models~2. We see that RS has positive effects for the three examined categories of technological impact, regardless of window length, and the effect is the most prominent for likelihood of patent citations. Figure~\ref{fig:coeff}B shows effect sizes of variety, as quantified as the changes of technological impact for citing one more field in the full models where the three aspects of IDR are all considered together (Models~6). It indicates that variety has the largest effect for likelihood of patent citations and the smallest effect for number of patent citations. Figure~\ref{fig:coeff}C suggests that balance has a consistently positive effect across the three groups of technological impact and across window lengths, and one standard deviation increase in balance has the largest effects on likelihood of attaining technological impact. Figure~\ref{fig:coeff}D illustrates mixed effects of disparity on different operationalizations of technological impact: While it has a negative effect on likelihood of achieving technological impact, it positively correlates with weighted number of patent citations. Note that two caveats remain when interpreting these results: One is that when studying likelihood of patent citations, the number of observations for 15-year citation window is smaller than the observations for 5- and 10-year windows, and the other is that, when studying (weighted) number of patent citations, the number of observations for the three citation window lengths are different, as a wider window will make more papers to get cited and consequently be included in regression analysis.

Second, one may argue that scientific citations---the number of scientific articles that cite the focal paper---may be a confounding variable, considering that the number of scientific citations may signify the quality of a paper, which may affect its chance to be diffused to the technology space. Previous studies have indeed found that scientific citations correlate with both patent citations \citep{Ke-compare-2018} and interdisciplinarity \citep{Wang-impact-2015, Yegros-interdisc-2015}. To address this concern, in Tables~S1--S9 in the supplementary materials, we include scientific citations as a control variable in all regression models, and the results remain similar. Note that we count scientific citations accrued in 5, 10, and 15 year, similar to patent citations.

Third, we have presented modeling results without including the indicator of whether a paper involves international collaboration as a control variable. This is because our corpus covers papers published in a long period of time (23 years) and for a significant portion of them (19.5\%; Table~\ref{tab:var}), we lack enough affiliation information to allow us to calculate this variable. In Tables~S10--S18 in the supplementary materials, we present the modeling results considering international collaboration, where the observations are the subset of papers with the international collaboration variable available. We make two observations. First, the associations between technological impact and the RS index and the three IDR indicators remain robust. Second, interestingly, international collaboration is negatively correlated with both the likelihood of getting patent citations and the unweighted and weighted number of patent citations, regardless of citation-window length. This means that for comparable papers in the same field and year, papers resulted from international collaboration are less likely to get patent citations than papers involving only domestic collaboration. Conditional on receiving citations from patents, international collaboration papers are cited less than those resulted from domestic collaboration. These associations are robustness even after controlling for scientific impact (Models~11 in Tables~S10--S18), and go in the opposite direction from the positive linkage between international collaboration and scientific impact.

Fourth, if a paper is associated with multiple SCs, we have picked the SC that is most represented in its reference list. Here we also employ another procedure, which is to create a dummy variable for each field and then to assign the value of one to the dummy variables corresponding to the SCs associated with a paper. In Tables~S19--S27 in the supplementary materials, we present the modeling results using this procedure, and our conclusions still hold.

Finally, we have restricted the sample to papers with patent citations when studying intensity of technological impact. To address the selection bias concern, in Tables~S28--S33 in the supplementary materials, we show that our results are robust if we use the Heckman two-step technique.

\section{Discussion} \label{sec:dis}

The main purpose of this work is to explore the relationship between IDR and technological impact. Our primary motivation is that while there has been many studies showing academic impact of IDR, its societal benefits remain under-explored, despite a major argument for promoting IDR is its potential to fulfill societal needs. In this work, we focus on one aspect of societal benefits, that is contributing to the development of patented technologies, as measured as getting cited by patents and hence achieving technological impact.

We have followed the literature that defines IDR from the knowledge integration perspective and relies on the disciplinary information of references cited by a focal paper. However, there has been a lack of consensus on a single ``best'' indicator for IDR \citep{Wang-consistency-2020}. We thus operationalize the extent of IDR using three popular indicators, namely variety, balance, and disparity, and an integrated metric called the Rao-Stirling index. They allow us to test the effects of different dimensions of IDR on technological impact. We have introduced three groups of indicators to capture technological impact, all of which are based on citations received from patents. These indicators are: (1) whether a paper has been cited by patents; (2) the number of patent citations; and (3) the number of patent citations weighted by the impact of those citing patents themselves. Using regression techniques, we find that RS has positive effects on all the three categories of technological impact, regardless of citation window length. Variety and balance have positive effects on the likelihood of being cited by patents, whereas disparity has a negative effect. These results are robust to different citation window lengths and indicate that for papers published in the same year and in the same field, those that cite more fields and whose distributions over cited fields are more even enjoy higher chances to achieve technological impact, but both positive effects can be offset if papers draw upon distant fields. These linkages are consistent with the view that drawing on knowledge from multiple fields is beneficial, so is drawing on an array of fields that are relatively close to each other, in accordance with existing research on the relationship between interdisciplinarity and scientific impact \citep{Yegros-interdisc-2015}. Yet, one may still note the differences in the samples of articles analyzed: Our analysis was performed only on biomedical papers, whereas papers from a wider spectrum of disciplines were included in previous studies.

We further find that conditional on getting cited by patents, variety is negatively associated with both raw and quality-adjusted number of patent citations, whereas disparity has a positive effect. These associations contrast themselves with the associations with likelihood of patent citations. While the effect sizes for variety is rather small, the relatively large effect sizes for disparity warrant discussions. We posit that the contrasting associations between disparity and likelihood and number of patent citations could be due to two factors at play. First, although the knowledge recombination theory implicates that studies committed to a recombinant search process is important for innovations, distal IDR might be perceived as too risky or put too much cognitive burden to technological audiences, which may hinder its use in technological development. Second, once diffused into the technology space, distal IDR, which integrates dissimilar bodies of knowledge and epistemic approaches in diverse disciplines, may bear applicability to more diverse technological fields and have greater tendency for connecting upstream research with downstream applications and for recognizing their market potential, thereby contributing to the development of science-based inventions.

Our work contributes to three lines of literature. First, our work enriches the multi-facet nature of the notion of impact. Traditionally, the impact of research has overwhelmingly implied as impact upon the academia, \emph{i.e.}, scientific impact. However, science funding agencies have increasingly emphasized impact of scientific research beyond the academia, \emph{i.e.}, the so-called ``broader impact''. This has prompted a field called altmetrics (alternative metrics) that attempts to search for impact in diverse domains other than science, such as various social media platforms and policy documents. Yet, social media mentions are transient, and whether they capture broader impact remains to be seen, since diverse stakeholders may participate in social media conversations \citep{Ke2017twitter}. Our work contributes to this area by shifting the attention to patent documents and expanding the conception of impact to the technological dimension. On a related note, we extend empirical characterizations of the benefits of IDR from the dominantly studied academic ones to technological ones.

Second, our work contributes to the sparse, yet emerging literature on the relationship between interdisciplinarity and university-industry interactions, considering that the vast majority of cited science is contributed by universities and companies are dominant assignees of citing patents \citep{Ke-analysis-2020}. Previous studies in this stream have argued that scientists whose research spans multiple disciplines are well positioned to engage in university-industry interactions \citep{Giuliani2010who, Rijnsoever2008resource}. Empirical evidence has also suggested the relationship between scientist's IDR orientation and their interaction with industry \citep{Este-relation-2019}. Our work focuses instead on individual research works, offering both theoretical argument and empirical scrutiny about IDR and its industry relevance.

Third, our work adds to the literature about the contribution of science to technology. The question of how scientific research can contribute to technological development has long been an important one in innovation studies. Extant literature has answered this question from the perspectives of institutional factors facilitating technology transfer \citep{Debackere-role-2005, Gregorio-why-2003, Cockburn-public-1996, Gambardella-competitive-1992} and scientific capacity of technologies. This work explores intrinsic characteristics of science itself, aligning with other previous works \citep{Ke-tech-2020}.

Our work may also carry implications for research policy. First, given the persistently strong associations between IDR and technological impact, policy encouragement of continuous fostering and supporting IDR seems important. These may include providing interdisciplinary training to early-career researchers, funding programs supporting cross-disciplinary collaborations, creating a collegial atmosphere that supports researchers with interdisciplinary background, and establishing better promotion criteria suitable for career progression of IDR researchers.

Second, the differentiated relationships between different aspects of IDR and technological impact revealed from our study are not entirely the same as the effects of IDR on scientific impact found in previous studies \citep{Yegros-interdisc-2015, Wang-impact-2015}. This may suggest more sophisticated policies. Specifically, one one hand, our results indicate a positive effect of variety on technological impact, which resonates with previous works that found variety is positively associated with scientific impact. From this perspective, policymakers may encourage cross-disciplinary research. On the other hand, the positive relationship between balance and technological impact contrasts with the negative effect of balance on scientific impact. The negative linkage between balance and scientific impact indicates that one effective consideration to yield scientific impact is to root research in one discipline and in the meantime, source from diverse other disciplines. Such a strategy, however, may be less effective to generate technological impact, which requires drawing knowledge from different disciplines evenly.

Some limitations remain in our work that may serve as future research. First, we have focused on the biomedicine domain. Future work could extend our analyses to other research areas like physical sciences to examine the generalizability of our findings. Second, we have looked at front-page citations to scientific papers, ignoring citations in the full-text of patent documents. This may underestimate the science-technology linkage. Further work can explore whether and how our results may differ if technological impact is quantified using full-text patent citations. In this regard, some recent studies have compared front-page citations with full-text ones, finding that papers that are cited in full-text are also more likely to be cited in the front-page of the same patent if these papers feature certain characteristics such as moderately basic and less interdisciplinary \citep{wang2021two}. Third, we have considered only one characteristic of citing patents---their impact---when constructing technological impact measures of papers. Followup studies could assess other characteristics of citing patents like novelty and disruptiveness. Moreover, similar to existing literature that constructs a paper citation network and then uses PageRank and other centrality measures to quantify scientific impact of papers, one may build an integrated citation network of science and technology and study centralities of nodes (papers and patents) in the network, which would provide quantification of impact that captures higher order information. Fourth, due to the data limitation, we have examined papers published until 2002, while funding programs may be interested in more recent science. Future work may include more recent papers if data about their patent citations are available. Still, our work not only is the first to examine the linkage between interdisciplinary research and technological impact but also does so in a large-scale. Despite these limitations, this work furthers our understanding of the interaction between interdisciplinary science and technology.

\section*{Acknowledgments}

Part of this work was performed while the author was with Northeastern University and Syracuse University. I acknowledge the data and computing resources provided there. The work described in this paper was partially supported by the National Natural Science Foundation of China (72204206), City University of Hong Kong (Project No. 9610552), and Hong Kong Institute for Data Science.

\clearpage
\appendix

\setcounter{table}{0}
\makeatletter
\renewcommand{\thetable}{A\@arabic\c@table}
\makeatother

\begin{landscape}
\begin{table*}[t!]
\centering
\caption{Correlations between variables.}
\label{tab:corr}
\begin{tabular}{l  c  c  c  c  c  c  c  c  c  c  c  c  c }
\hline
&C5&C10&C15&JIF&MeSH&Authors&Variety&Balance&Disparity&RS&Cited5&Cited10&Cited15\\ \hline
C5&1.000\\
C10&0.935&1.000\\
C15&0.841&0.969&1.000\\
JIF&0.502&0.426&0.367&1.000\\
MeSH&0.191&0.165&0.142&0.234&1.000\\
Authors&0.175&0.151&0.131&0.139&0.258&1.000\\
Variety&0.200&0.181&0.161&0.294&0.331&0.188&1.000\\
Balance&0.082&0.073&0.065&0.127&0.133&0.120&0.509&1.000\\
Disparity&0.017&0.033&0.041&-0.027&0.066&0.094&0.454&0.628&1.000\\
RS&-0.008&0.011&0.022&-0.073&0.043&0.097&0.543&0.678&0.834&1.000\\
Cited5&0.235&0.206&0.179&0.179&0.065&0.078&0.083&0.037&0.006&0.006&1.000\\
Cited10&0.293&0.263&0.231&0.258&0.108&0.111&0.141&0.059&0.011&0.009&0.612&1.000\\
Cited15&0.301&0.271&0.240&0.274&0.116&0.115&0.156&0.066&0.013&0.011&0.532&0.868&1.000\\
\hline
\end{tabular}
\end{table*}
\end{landscape}

\begin{landscape}
\begin{table*}[t!]
\centering
\caption{Logistic regression modeling of whether a paper has patent citations in 5 years. \label{tab:citedbypat5}}
\begin{tabular}{l*{10}{c}}
\toprule%
& (1) & (2) & (3) & (4) & (5) & (6) & (7) & (8) & (9) & (10) \\
\midrule
JIF                 &       0.117\sym{***}&       0.121\sym{***}&       0.115\sym{***}&       0.117\sym{***}&       0.118\sym{***}&       0.114\sym{***}&       0.121\sym{***}&       0.112\sym{***}&       0.112\sym{***}&       0.114\sym{***}\\
                    &  (0.000760)         &  (0.000769)         &  (0.000762)         &  (0.000760)         &  (0.000765)         &  (0.000766)         &  (0.000772)         &  (0.000768)         &  (0.000770)         &  (0.000767)         \\
MeSH                &      0.0382\sym{***}&      0.0400\sym{***}&      0.0294\sym{***}&      0.0377\sym{***}&      0.0391\sym{***}&      0.0281\sym{***}&      0.0394\sym{***}&      0.0262\sym{***}&      0.0267\sym{***}&      0.0282\sym{***}\\
                    &  (0.000576)         &  (0.000577)         &  (0.000588)         &  (0.000577)         &  (0.000577)         &  (0.000596)         &  (0.000579)         &  (0.000600)         &  (0.000602)         &  (0.000598)         \\
Authors ($\ln$)     &       0.508\sym{***}&       0.485\sym{***}&       0.464\sym{***}&       0.489\sym{***}&       0.497\sym{***}&       0.463\sym{***}&       0.478\sym{***}&       0.459\sym{***}&       0.462\sym{***}&       0.464\sym{***}\\
                    &   (0.00533)         &   (0.00536)         &   (0.00539)         &   (0.00536)         &   (0.00536)         &   (0.00540)         &   (0.00538)         &   (0.00541)         &   (0.00541)         &   (0.00541)         \\
RS                  &                     &       1.010\sym{***}&                     &                     &                     &                     &       2.080\sym{***}&                     &                     &                     \\
                    &                     &    (0.0236)         &                     &                     &                     &                     &    (0.0753)         &                     &                     &                     \\
Variety             &                     &                     &      0.0778\sym{***}&                     &                     &      0.0785\sym{***}&                     &       0.171\sym{***}&      0.0781\sym{***}&      0.0789\sym{***}\\
                    &                     &                     &  (0.000892)         &                     &                     &   (0.00100)         &                     &   (0.00353)         &   (0.00101)         &   (0.00102)         \\
Balance             &                     &                     &                     &       0.718\sym{***}&                     &       0.433\sym{***}&                     &       0.268\sym{***}&       1.458\sym{***}&       0.450\sym{***}\\
                    &                     &                     &                     &    (0.0164)         &                     &    (0.0182)         &                     &    (0.0195)         &    (0.0652)         &    (0.0196)         \\
Disparity           &                     &                     &                     &                     &       0.448\sym{***}&      -0.442\sym{***}&                     &      -0.544\sym{***}&      -0.574\sym{***}&      -0.615\sym{***}\\
                    &                     &                     &                     &                     &    (0.0189)         &    (0.0229)         &                     &    (0.0234)         &    (0.0243)         &    (0.0767)         \\
$\text{RS}^2$       &                     &                     &                     &                     &                     &                     &      -2.083\sym{***}&                     &                     &                     \\
                    &                     &                     &                     &                     &                     &                     &     (0.139)         &                     &                     &                     \\
$\text{Variety}^2$  &                     &                     &                     &                     &                     &                     &                     &    -0.00528\sym{***}&                     &                     \\
                    &                     &                     &                     &                     &                     &                     &                     &  (0.000194)         &                     &                     \\
$\text{Balance}^2$  &                     &                     &                     &                     &                     &                     &                     &                     &      -0.889\sym{***}&                     \\
                    &                     &                     &                     &                     &                     &                     &                     &                     &    (0.0539)         &                     \\
$\text{Disparity}^2$&                     &                     &                     &                     &                     &                     &                     &                     &                     &       0.227\sym{*}  \\
                    &                     &                     &                     &                     &                     &                     &                     &                     &                     &    (0.0961)         \\
Constant            &      -4.578\sym{***}&      -4.950\sym{***}&      -5.184\sym{***}&      -5.132\sym{***}&      -4.784\sym{***}&      -5.322\sym{***}&      -4.999\sym{***}&      -5.343\sym{***}&      -5.443\sym{***}&      -5.317\sym{***}\\
                    &     (0.204)         &     (0.204)         &     (0.205)         &     (0.204)         &     (0.204)         &     (0.205)         &     (0.204)         &     (0.204)         &     (0.205)         &     (0.205)         \\
\hline
Field fe            & $\checkmark$        & $\checkmark$        & $\checkmark$        & $\checkmark$        & $\checkmark$        & $\checkmark$        & $\checkmark$        & $\checkmark$        & $\checkmark$        & $\checkmark$ \\
Year fe             & $\checkmark$        & $\checkmark$        & $\checkmark$        & $\checkmark$        & $\checkmark$        & $\checkmark$        & $\checkmark$        & $\checkmark$        & $\checkmark$        & $\checkmark$ \\
\(N\)               &     5445517         &     5445517         &     5445517         &     5445517         &     5445517         &     5445517         &     5445517         &     5445517         &     5445517         &     5445517         \\
Pseudo \(R^{2}\)    &       0.141         &       0.142         &       0.146         &       0.143         &       0.141         &       0.147         &       0.142         &       0.147         &       0.147         &       0.147         \\
\textit{BIC}        &   1180078.6         &   1178269.5         &   1172729.0         &   1178000.5         &   1179526.5         &   1172016.8         &   1178055.8         &   1171237.9         &   1171755.4         &   1172026.7         \\
\bottomrule
\multicolumn{5}{l}{\footnotesize Standard errors in parentheses}\\
\multicolumn{5}{l}{\footnotesize \sym{*} \(p<0.05\), \sym{**} \(p<0.01\), \sym{***} \(p<0.001\)}\\
\end{tabular}
\end{table*}
\end{landscape}

\begin{landscape}
\begin{table*}[t!]
\centering
\caption{Logistic regression modeling of whether a paper has patent citations in 15 years. \label{tab:citedbypat15}}
\begin{tabular}{l*{10}{c}}
\toprule%
& (1) & (2) & (3) & (4) & (5) & (6) & (7) & (8) & (9) & (10) \\
\midrule
JIF                 &       0.175\sym{***}&       0.182\sym{***}&       0.167\sym{***}&       0.175\sym{***}&       0.178\sym{***}&       0.166\sym{***}&       0.183\sym{***}&       0.162\sym{***}&       0.162\sym{***}&       0.165\sym{***}\\
                    &  (0.000809)         &  (0.000820)         &  (0.000806)         &  (0.000810)         &  (0.000814)         &  (0.000813)         &  (0.000821)         &  (0.000816)         &  (0.000820)         &  (0.000813)         \\
MeSH                &      0.0465\sym{***}&      0.0479\sym{***}&      0.0359\sym{***}&      0.0455\sym{***}&      0.0471\sym{***}&      0.0350\sym{***}&      0.0470\sym{***}&      0.0326\sym{***}&      0.0332\sym{***}&      0.0352\sym{***}\\
                    &  (0.000460)         &  (0.000461)         &  (0.000469)         &  (0.000461)         &  (0.000460)         &  (0.000473)         &  (0.000463)         &  (0.000476)         &  (0.000478)         &  (0.000475)         \\
Authors ($\ln$)     &       0.457\sym{***}&       0.431\sym{***}&       0.414\sym{***}&       0.434\sym{***}&       0.442\sym{***}&       0.412\sym{***}&       0.425\sym{***}&       0.407\sym{***}&       0.411\sym{***}&       0.413\sym{***}\\
                    &   (0.00383)         &   (0.00385)         &   (0.00387)         &   (0.00385)         &   (0.00385)         &   (0.00388)         &   (0.00386)         &   (0.00389)         &   (0.00388)         &   (0.00388)         \\
RS                  &                     &       1.151\sym{***}&                     &                     &                     &                     &       2.295\sym{***}&                     &                     &                     \\
                    &                     &    (0.0168)         &                     &                     &                     &                     &    (0.0524)         &                     &                     &                     \\
Variety             &                     &                     &      0.0879\sym{***}&                     &                     &      0.0866\sym{***}&                     &       0.198\sym{***}&      0.0858\sym{***}&      0.0873\sym{***}\\
                    &                     &                     &  (0.000666)         &                     &                     &  (0.000759)         &                     &   (0.00266)         &  (0.000762)         &  (0.000771)         \\
Balance             &                     &                     &                     &       0.749\sym{***}&                     &       0.414\sym{***}&                     &       0.206\sym{***}&       1.498\sym{***}&       0.441\sym{***}\\
                    &                     &                     &                     &    (0.0106)         &                     &    (0.0122)         &                     &    (0.0133)         &    (0.0429)         &    (0.0133)         \\
Disparity           &                     &                     &                     &                     &       0.583\sym{***}&      -0.390\sym{***}&                     &      -0.508\sym{***}&      -0.550\sym{***}&      -0.645\sym{***}\\
                    &                     &                     &                     &                     &    (0.0130)         &    (0.0161)         &                     &    (0.0165)         &    (0.0172)         &    (0.0536)         \\
$\text{RS}^2$       &                     &                     &                     &                     &                     &                     &      -2.330\sym{***}&                     &                     &                     \\
                    &                     &                     &                     &                     &                     &                     &     (0.101)         &                     &                     &                     \\
$\text{Variety}^2$  &                     &                     &                     &                     &                     &                     &                     &    -0.00680\sym{***}&                     &                     \\
                    &                     &                     &                     &                     &                     &                     &                     &  (0.000156)         &                     &                     \\
$\text{Balance}^2$  &                     &                     &                     &                     &                     &                     &                     &                     &      -0.965\sym{***}&                     \\
                    &                     &                     &                     &                     &                     &                     &                     &                     &    (0.0364)         &                     \\
$\text{Disparity}^2$&                     &                     &                     &                     &                     &                     &                     &                     &                     &       0.340\sym{***}\\
                    &                     &                     &                     &                     &                     &                     &                     &                     &                     &    (0.0683)         \\
Constant            &      -5.523\sym{***}&      -5.840\sym{***}&      -5.934\sym{***}&      -6.110\sym{***}&      -5.757\sym{***}&      -6.094\sym{***}&      -5.941\sym{***}&      -6.217\sym{***}&      -6.211\sym{***}&      -6.081\sym{***}\\
                    &    (0.0867)         &    (0.0868)         &    (0.0869)         &    (0.0871)         &    (0.0868)         &    (0.0873)         &    (0.0869)         &    (0.0873)         &    (0.0875)         &    (0.0873)         \\
\hline
Field fe            & $\checkmark$        & $\checkmark$        & $\checkmark$        & $\checkmark$        & $\checkmark$        & $\checkmark$        & $\checkmark$        & $\checkmark$        & $\checkmark$        & $\checkmark$        \\
Year fe             & $\checkmark$        & $\checkmark$        & $\checkmark$        & $\checkmark$        & $\checkmark$        & $\checkmark$        & $\checkmark$        & $\checkmark$        & $\checkmark$        & $\checkmark$        \\
\(N\)               &     3948327         &     3948327         &     3948327         &     3948327         &     3948327         &     3948327         &     3948327         &     3948327         &     3948327         &     3948327         \\
Pseudo \(R^{2}\)    &       0.169         &       0.171         &       0.176         &       0.171         &       0.170         &       0.177         &       0.171         &       0.178         &       0.177         &       0.177         \\
\textit{BIC}        &   1904897.6         &   1900196.9         &   1887781.7         &   1899436.6         &   1902861.7         &   1886469.4         &   1899669.4         &   1884479.8         &   1885774.1         &   1886459.9         \\
\bottomrule
\multicolumn{5}{l}{\footnotesize Standard errors in parentheses}\\
\multicolumn{5}{l}{\footnotesize \sym{*} \(p<0.05\), \sym{**} \(p<0.01\), \sym{***} \(p<0.001\)}\\
\end{tabular}
\end{table*}
\end{landscape}

\begin{landscape}
\begin{table*}[t!]
\centering
\caption{Negative binomial regression modeling of 5-year patent citations. \label{tab:patc5}}
\begin{tabular}{l*{10}{c}}
\toprule%
& (1) & (2) & (3) & (4) & (5) & (6) & (7) & (8) & (9) & (10) \\
\midrule
JIF                 &      0.0175\sym{***}&      0.0179\sym{***}&      0.0175\sym{***}&      0.0175\sym{***}&      0.0178\sym{***}&      0.0178\sym{***}&      0.0178\sym{***}&      0.0179\sym{***}&      0.0181\sym{***}&      0.0178\sym{***}\\
                    &  (0.000542)         &  (0.000545)         &  (0.000542)         &  (0.000542)         &  (0.000544)         &  (0.000544)         &  (0.000546)         &  (0.000545)         &  (0.000546)         &  (0.000545)         \\
MeSH                &    -0.00258\sym{***}&    -0.00222\sym{***}&    -0.00266\sym{***}&    -0.00259\sym{***}&    -0.00229\sym{***}&    -0.00215\sym{***}&    -0.00219\sym{***}&    -0.00199\sym{***}&    -0.00180\sym{***}&    -0.00213\sym{***}\\
                    &  (0.000472)         &  (0.000475)         &  (0.000477)         &  (0.000472)         &  (0.000474)         &  (0.000485)         &  (0.000475)         &  (0.000487)         &  (0.000488)         &  (0.000486)         \\
Authors ($\ln$)     &      0.0982\sym{***}&      0.0955\sym{***}&      0.0978\sym{***}&      0.0969\sym{***}&      0.0964\sym{***}&      0.0964\sym{***}&      0.0959\sym{***}&      0.0968\sym{***}&      0.0961\sym{***}&      0.0965\sym{***}\\
                    &   (0.00426)         &   (0.00428)         &   (0.00428)         &   (0.00426)         &   (0.00427)         &   (0.00428)         &   (0.00429)         &   (0.00428)         &   (0.00428)         &   (0.00429)         \\
RS                  &                     &       0.135\sym{***}&                     &                     &                     &                     &      0.0612         &                     &                     &                     \\
                    &                     &    (0.0206)         &                     &                     &                     &                     &    (0.0647)         &                     &                     &                     \\
Variety             &                     &                     &    0.000867         &                     &                     &    -0.00181\sym{*}  &                     &     -0.0115\sym{***}&    -0.00171\sym{*}  &    -0.00173         \\
                    &                     &                     &  (0.000780)         &                     &                     &  (0.000868)         &                     &   (0.00294)         &  (0.000867)         &  (0.000882)         \\
Balance             &                     &                     &                     &      0.0837\sym{***}&                     &      0.0684\sym{***}&                     &      0.0845\sym{***}&      -0.261\sym{***}&      0.0709\sym{***}\\
                    &                     &                     &                     &    (0.0155)         &                     &    (0.0166)         &                     &    (0.0173)         &    (0.0580)         &    (0.0173)         \\
Disparity           &                     &                     &                     &                     &      0.0992\sym{***}&      0.0919\sym{***}&                     &       0.101\sym{***}&       0.131\sym{***}&      0.0584         \\
                    &                     &                     &                     &                     &    (0.0171)         &    (0.0192)         &                     &    (0.0194)         &    (0.0203)         &    (0.0646)         \\
$\text{RS}^2$       &                     &                     &                     &                     &                     &                     &       0.140         &                     &                     &                     \\
                    &                     &                     &                     &                     &                     &                     &     (0.116)         &                     &                     &                     \\
$\text{Variety}^2$  &                     &                     &                     &                     &                     &                     &                     &    0.000546\sym{***}&                     &                     \\
                    &                     &                     &                     &                     &                     &                     &                     &  (0.000158)         &                     &                     \\
$\text{Balance}^2$  &                     &                     &                     &                     &                     &                     &                     &                     &       0.280\sym{***}&                     \\
                    &                     &                     &                     &                     &                     &                     &                     &                     &    (0.0472)         &                     \\
$\text{Disparity}^2$&                     &                     &                     &                     &                     &                     &                     &                     &                     &      0.0441         \\
                    &                     &                     &                     &                     &                     &                     &                     &                     &                     &    (0.0811)         \\
Constant            &       0.627         &       0.576         &       0.623         &       0.561         &       0.571         &       0.530         &       0.583         &       0.549         &       0.576         &       0.532         \\
                    &     (0.829)         &     (0.829)         &     (0.829)         &     (0.829)         &     (0.829)         &     (0.829)         &     (0.829)         &     (0.829)         &     (0.829)         &     (0.829)         \\
\hline
lnalpha             &      -1.683\sym{***}&      -1.683\sym{***}&      -1.683\sym{***}&      -1.683\sym{***}&      -1.683\sym{***}&      -1.683\sym{***}&      -1.683\sym{***}&      -1.683\sym{***}&      -1.684\sym{***}&      -1.683\sym{***}\\
                    &    (0.0103)         &    (0.0103)         &    (0.0103)         &    (0.0103)         &    (0.0103)         &    (0.0103)         &    (0.0103)         &    (0.0103)         &    (0.0103)         &    (0.0103)         \\
\hline
Field fe            & $\checkmark$        & $\checkmark$        & $\checkmark$        & $\checkmark$        & $\checkmark$        & $\checkmark$        & $\checkmark$        & $\checkmark$        & $\checkmark$        & $\checkmark$        \\
Year fe             & $\checkmark$        & $\checkmark$        & $\checkmark$        & $\checkmark$        & $\checkmark$        & $\checkmark$        & $\checkmark$        & $\checkmark$        & $\checkmark$        & $\checkmark$        \\
\(N\)               &      149643         &      149643         &      149643         &      149643         &      149643         &      149643         &      149643         &      149643         &      149643         &      149643         \\
\textit{BIC}        &    490799.0         &    490768.2         &    490809.7         &    490781.7         &    490777.4         &    490782.5         &    490778.7         &    490782.5         &    490759.4         &    490794.1         \\
\bottomrule
\multicolumn{5}{l}{\footnotesize Standard errors in parentheses}\\
\multicolumn{5}{l}{\footnotesize \sym{*} \(p<0.05\), \sym{**} \(p<0.01\), \sym{***} \(p<0.001\)}\\
\end{tabular}
\end{table*}
\end{landscape}

\begin{landscape}
\begin{table*}[t!]
\centering
\caption{Negative binomial regression modeling of 15-year patent citations. \label{tab:patc15}}
\begin{tabular}{l*{10}{c}}
\toprule%
& (1) & (2) & (3) & (4) & (5) & (6) & (7) & (8) & (9) & (10) \\
\midrule
JIF                 &      0.0399\sym{***}&      0.0402\sym{***}&      0.0397\sym{***}&      0.0400\sym{***}&      0.0401\sym{***}&      0.0397\sym{***}&      0.0402\sym{***}&      0.0398\sym{***}&      0.0399\sym{***}&      0.0396\sym{***}\\
                    &  (0.000526)         &  (0.000530)         &  (0.000525)         &  (0.000526)         &  (0.000528)         &  (0.000529)         &  (0.000531)         &  (0.000531)         &  (0.000533)         &  (0.000530)         \\
MeSH                &     0.00108\sym{**} &     0.00127\sym{**} &    0.000484         &     0.00107\sym{**} &     0.00120\sym{**} &    0.000503         &     0.00124\sym{**} &    0.000590         &    0.000701         &    0.000702         \\
                    &  (0.000408)         &  (0.000410)         &  (0.000413)         &  (0.000408)         &  (0.000410)         &  (0.000418)         &  (0.000411)         &  (0.000420)         &  (0.000421)         &  (0.000419)         \\
Authors ($\ln$)     &       0.137\sym{***}&       0.136\sym{***}&       0.134\sym{***}&       0.135\sym{***}&       0.136\sym{***}&       0.134\sym{***}&       0.136\sym{***}&       0.134\sym{***}&       0.133\sym{***}&       0.135\sym{***}\\
                    &   (0.00336)         &   (0.00338)         &   (0.00338)         &   (0.00337)         &   (0.00337)         &   (0.00338)         &   (0.00339)         &   (0.00338)         &   (0.00338)         &   (0.00339)         \\
RS                  &                     &      0.0823\sym{***}&                     &                     &                     &                     &       0.130\sym{**} &                     &                     &                     \\
                    &                     &    (0.0165)         &                     &                     &                     &                     &    (0.0500)         &                     &                     &                     \\
Variety             &                     &                     &     0.00617\sym{***}&                     &                     &     0.00490\sym{***}&                     &   -0.000248         &     0.00503\sym{***}&     0.00584\sym{***}\\
                    &                     &                     &  (0.000640)         &                     &                     &  (0.000719)         &                     &   (0.00248)         &  (0.000720)         &  (0.000734)         \\
Balance             &                     &                     &                     &       0.130\sym{***}&                     &       0.115\sym{***}&                     &       0.124\sym{***}&     -0.0464         &       0.139\sym{***}\\
                    &                     &                     &                     &    (0.0113)         &                     &    (0.0123)         &                     &    (0.0131)         &    (0.0425)         &    (0.0129)         \\
Disparity           &                     &                     &                     &                     &      0.0500\sym{***}&     -0.0395\sym{**} &                     &     -0.0347\sym{*}  &     -0.0185         &      -0.333\sym{***}\\
                    &                     &                     &                     &                     &    (0.0131)         &    (0.0149)         &                     &    (0.0150)         &    (0.0158)         &    (0.0490)         \\
$\text{RS}^2$       &                     &                     &                     &                     &                     &                     &     -0.0955         &                     &                     &                     \\
                    &                     &                     &                     &                     &                     &                     &    (0.0940)         &                     &                     &                     \\
$\text{Variety}^2$  &                     &                     &                     &                     &                     &                     &                     &    0.000307\sym{*}  &                     &                     \\
                    &                     &                     &                     &                     &                     &                     &                     &  (0.000142)         &                     &                     \\
$\text{Balance}^2$  &                     &                     &                     &                     &                     &                     &                     &                     &       0.141\sym{***}&                     \\
                    &                     &                     &                     &                     &                     &                     &                     &                     &    (0.0354)         &                     \\
$\text{Disparity}^2$&                     &                     &                     &                     &                     &                     &                     &                     &                     &       0.397\sym{***}\\
                    &                     &                     &                     &                     &                     &                     &                     &                     &                     &    (0.0630)         \\
Constant            &      0.0871         &      0.0533         &      0.0643         &     -0.0279         &      0.0572         &    -0.00885         &      0.0499         &    -0.00372         &     0.00409         &    -0.00248         \\
                    &     (0.651)         &     (0.651)         &     (0.651)         &     (0.651)         &     (0.651)         &     (0.651)         &     (0.651)         &     (0.651)         &     (0.651)         &     (0.651)         \\
\hline
lnalpha             &      -0.264\sym{***}&      -0.264\sym{***}&      -0.264\sym{***}&      -0.264\sym{***}&      -0.264\sym{***}&      -0.265\sym{***}&      -0.264\sym{***}&      -0.265\sym{***}&      -0.265\sym{***}&      -0.265\sym{***}\\
                    &   (0.00294)         &   (0.00294)         &   (0.00294)         &   (0.00294)         &   (0.00294)         &   (0.00294)         &   (0.00294)         &   (0.00294)         &   (0.00294)         &   (0.00294)         \\
\hline
Field fe            & $\checkmark$        & $\checkmark$        & $\checkmark$        & $\checkmark$        & $\checkmark$        & $\checkmark$        & $\checkmark$        & $\checkmark$        & $\checkmark$        & $\checkmark$        \\
Year fe             & $\checkmark$        & $\checkmark$        & $\checkmark$        & $\checkmark$        & $\checkmark$        & $\checkmark$        & $\checkmark$        & $\checkmark$        & $\checkmark$        & $\checkmark$        \\
\(N\)               &      334124         &      334124         &      334124         &      334124         &      334124         &      334124         &      334124         &      334124         &      334124         &      334124         \\
\textit{BIC}        &   1624035.0         &   1624023.0         &   1623954.8         &   1623914.8         &   1624033.0         &   1623893.7         &   1624034.7         &   1623901.7         &   1623890.7         &   1623866.7         \\
\bottomrule
\multicolumn{5}{l}{\footnotesize Standard errors in parentheses}\\
\multicolumn{5}{l}{\footnotesize \sym{*} \(p<0.05\), \sym{**} \(p<0.01\), \sym{***} \(p<0.001\)}\\
\end{tabular}
\end{table*}
\end{landscape}

\begin{landscape}
\begin{table*}[t!]
\centering
\caption{Negative binomial regression modeling of weighted 5-year patent citations. \label{tab:wpatc5}}
\begin{tabular}{l*{10}{c}}
\toprule%
& (1) & (2) & (3) & (4) & (5) & (6) & (7) & (8) & (9) & (10) \\
\midrule
JIF                 &      0.0167\sym{***}&      0.0181\sym{***}&      0.0170\sym{***}&      0.0168\sym{***}&      0.0180\sym{***}&      0.0192\sym{***}&      0.0173\sym{***}&      0.0202\sym{***}&      0.0205\sym{***}&      0.0190\sym{***}\\
                    &   (0.00102)         &   (0.00103)         &   (0.00103)         &   (0.00102)         &   (0.00103)         &   (0.00104)         &   (0.00103)         &   (0.00104)         &   (0.00105)         &   (0.00104)         \\
MeSH                &     -0.0266\sym{***}&     -0.0251\sym{***}&     -0.0249\sym{***}&     -0.0266\sym{***}&     -0.0252\sym{***}&     -0.0214\sym{***}&     -0.0243\sym{***}&     -0.0202\sym{***}&     -0.0201\sym{***}&     -0.0205\sym{***}\\
                    &  (0.000797)         &  (0.000806)         &  (0.000808)         &  (0.000797)         &  (0.000803)         &  (0.000825)         &  (0.000810)         &  (0.000830)         &  (0.000833)         &  (0.000829)         \\
Authors ($\ln$)     &      0.0543\sym{***}&      0.0448\sym{***}&      0.0637\sym{***}&      0.0512\sym{***}&      0.0460\sym{***}&      0.0565\sym{***}&      0.0521\sym{***}&      0.0613\sym{***}&      0.0571\sym{***}&      0.0629\sym{***}\\
                    &   (0.00712)         &   (0.00714)         &   (0.00716)         &   (0.00713)         &   (0.00713)         &   (0.00716)         &   (0.00716)         &   (0.00717)         &   (0.00716)         &   (0.00717)         \\
RS                  &                     &       0.448\sym{***}&                     &                     &                     &                     &      -0.919\sym{***}&                     &                     &                     \\
                    &                     &    (0.0338)         &                     &                     &                     &                     &     (0.109)         &                     &                     &                     \\
Variety             &                     &                     &     -0.0168\sym{***}&                     &                     &     -0.0307\sym{***}&                     &      -0.102\sym{***}&     -0.0308\sym{***}&     -0.0280\sym{***}\\
                    &                     &                     &   (0.00130)         &                     &                     &   (0.00144)         &                     &   (0.00514)         &   (0.00145)         &   (0.00146)         \\
Balance             &                     &                     &                     &       0.168\sym{***}&                     &       0.140\sym{***}&                     &       0.271\sym{***}&      -0.963\sym{***}&       0.262\sym{***}\\
                    &                     &                     &                     &    (0.0249)         &                     &    (0.0268)         &                     &    (0.0286)         &    (0.0943)         &    (0.0287)         \\
Disparity           &                     &                     &                     &                     &       0.405\sym{***}&       0.620\sym{***}&                     &       0.697\sym{***}&       0.763\sym{***}&      -0.731\sym{***}\\
                    &                     &                     &                     &                     &    (0.0276)         &    (0.0319)         &                     &    (0.0325)         &    (0.0340)         &     (0.109)         \\
$\text{RS}^2$       &                     &                     &                     &                     &                     &                     &       2.586\sym{***}&                     &                     &                     \\
                    &                     &                     &                     &                     &                     &                     &     (0.196)         &                     &                     &                     \\
$\text{Variety}^2$  &                     &                     &                     &                     &                     &                     &                     &     0.00399\sym{***}&                     &                     \\
                    &                     &                     &                     &                     &                     &                     &                     &  (0.000278)         &                     &                     \\
$\text{Balance}^2$  &                     &                     &                     &                     &                     &                     &                     &                     &       0.958\sym{***}&                     \\
                    &                     &                     &                     &                     &                     &                     &                     &                     &    (0.0779)         &                     \\
$\text{Disparity}^2$&                     &                     &                     &                     &                     &                     &                     &                     &                     &       1.786\sym{***}\\
                    &                     &                     &                     &                     &                     &                     &                     &                     &                     &     (0.138)         \\
Constant            &       0.260         &      0.0831         &       0.349         &       0.124         &      0.0261         &     -0.0461         &       0.209         &      0.0805         &      0.0881         &    -0.00563         \\
                    &     (1.755)         &     (1.754)         &     (1.754)         &     (1.755)         &     (1.754)         &     (1.752)         &     (1.754)         &     (1.752)         &     (1.752)         &     (1.752)         \\
\hline
lnalpha             &       0.731\sym{***}&       0.729\sym{***}&       0.730\sym{***}&       0.730\sym{***}&       0.729\sym{***}&       0.726\sym{***}&       0.728\sym{***}&       0.725\sym{***}&       0.725\sym{***}&       0.725\sym{***}\\
                    &   (0.00362)         &   (0.00362)         &   (0.00362)         &   (0.00362)         &   (0.00362)         &   (0.00362)         &   (0.00362)         &   (0.00362)         &   (0.00362)         &   (0.00362)         \\
\hline
Field fe            & $\checkmark$        & $\checkmark$        & $\checkmark$        & $\checkmark$        & $\checkmark$        & $\checkmark$        & $\checkmark$        & $\checkmark$        & $\checkmark$        & $\checkmark$        \\
Year fe             & $\checkmark$        & $\checkmark$        & $\checkmark$        & $\checkmark$        & $\checkmark$        & $\checkmark$        & $\checkmark$        & $\checkmark$        & $\checkmark$        & $\checkmark$        \\
\(N\)               &      149643         &      149643         &      149643         &      149643         &      149643         &      149643         &      149643         &      149643         &      149643         &      149643         \\
\textit{BIC}        &   1025918.8         &   1025754.7         &   1025765.8         &   1025885.9         &   1025718.0         &   1025294.2         &   1025587.7         &   1025088.5         &   1025152.2         &   1025135.9         \\
\bottomrule
\multicolumn{5}{l}{\footnotesize Standard errors in parentheses}\\
\multicolumn{5}{l}{\footnotesize \sym{*} \(p<0.05\), \sym{**} \(p<0.01\), \sym{***} \(p<0.001\)}\\
\end{tabular}
\end{table*}
\end{landscape}

\begin{landscape}
\begin{table*}[t!]
\centering
\caption{Negative binomial regression modeling of weighted 15-year patent citations. \label{tab:wpatc15}}
\begin{tabular}{l*{10}{c}}
\toprule%
& (1) & (2) & (3) & (4) & (5) & (6) & (7) & (8) & (9) & (10) \\
\midrule
JIF                 &      0.0295\sym{***}&      0.0310\sym{***}&      0.0300\sym{***}&      0.0295\sym{***}&      0.0311\sym{***}&      0.0329\sym{***}&      0.0302\sym{***}&      0.0339\sym{***}&      0.0348\sym{***}&      0.0326\sym{***}\\
                    &  (0.000890)         &  (0.000900)         &  (0.000894)         &  (0.000890)         &  (0.000896)         &  (0.000908)         &  (0.000900)         &  (0.000914)         &  (0.000922)         &  (0.000909)         \\
MeSH                &     -0.0221\sym{***}&     -0.0211\sym{***}&     -0.0207\sym{***}&     -0.0221\sym{***}&     -0.0210\sym{***}&     -0.0175\sym{***}&     -0.0201\sym{***}&     -0.0162\sym{***}&     -0.0158\sym{***}&     -0.0157\sym{***}\\
                    &  (0.000665)         &  (0.000669)         &  (0.000675)         &  (0.000665)         &  (0.000668)         &  (0.000685)         &  (0.000672)         &  (0.000689)         &  (0.000691)         &  (0.000689)         \\
Authors ($\ln$)     &      0.0944\sym{***}&      0.0866\sym{***}&       0.101\sym{***}&      0.0929\sym{***}&      0.0859\sym{***}&      0.0955\sym{***}&      0.0932\sym{***}&      0.0993\sym{***}&      0.0953\sym{***}&       0.103\sym{***}\\
                    &   (0.00547)         &   (0.00549)         &   (0.00550)         &   (0.00548)         &   (0.00548)         &   (0.00550)         &   (0.00551)         &   (0.00551)         &   (0.00550)         &   (0.00550)         \\
RS                  &                     &       0.381\sym{***}&                     &                     &                     &                     &      -1.054\sym{***}&                     &                     &                     \\
                    &                     &    (0.0258)         &                     &                     &                     &                     &    (0.0806)         &                     &                     &                     \\
Variety             &                     &                     &     -0.0129\sym{***}&                     &                     &     -0.0266\sym{***}&                     &      -0.102\sym{***}&     -0.0259\sym{***}&     -0.0208\sym{***}\\
                    &                     &                     &   (0.00104)         &                     &                     &   (0.00116)         &                     &   (0.00413)         &   (0.00117)         &   (0.00119)         \\
Balance             &                     &                     &                     &      0.0764\sym{***}&                     &     0.00214         &                     &       0.151\sym{***}&      -1.217\sym{***}&       0.204\sym{***}\\
                    &                     &                     &                     &    (0.0176)         &                     &    (0.0194)         &                     &    (0.0211)         &    (0.0673)         &    (0.0210)         \\
Disparity           &                     &                     &                     &                     &       0.409\sym{***}&       0.630\sym{***}&                     &       0.704\sym{***}&       0.793\sym{***}&      -1.514\sym{***}\\
                    &                     &                     &                     &                     &    (0.0199)         &    (0.0235)         &                     &    (0.0239)         &    (0.0250)         &    (0.0803)         \\
$\text{RS}^2$       &                     &                     &                     &                     &                     &                     &       2.830\sym{***}&                     &                     &                     \\
                    &                     &                     &                     &                     &                     &                     &     (0.151)         &                     &                     &                     \\
$\text{Variety}^2$  &                     &                     &                     &                     &                     &                     &                     &     0.00449\sym{***}&                     &                     \\
                    &                     &                     &                     &                     &                     &                     &                     &  (0.000236)         &                     &                     \\
$\text{Balance}^2$  &                     &                     &                     &                     &                     &                     &                     &                     &       1.086\sym{***}&                     \\
                    &                     &                     &                     &                     &                     &                     &                     &                     &    (0.0569)         &                     \\
$\text{Disparity}^2$&                     &                     &                     &                     &                     &                     &                     &                     &                     &       2.874\sym{***}\\
                    &                     &                     &                     &                     &                     &                     &                     &                     &                     &     (0.103)         \\
Constant            &       1.973\sym{*}  &       1.812         &       2.017\sym{*}  &       1.904         &       1.725         &       1.681         &       1.918\sym{*}  &       1.745         &       1.755         &       1.717         \\
                    &     (0.976)         &     (0.976)         &     (0.976)         &     (0.976)         &     (0.975)         &     (0.975)         &     (0.975)         &     (0.975)         &     (0.975)         &     (0.974)         \\
\hline
lnalpha             &       1.017\sym{***}&       1.016\sym{***}&       1.016\sym{***}&       1.017\sym{***}&       1.015\sym{***}&       1.014\sym{***}&       1.015\sym{***}&       1.013\sym{***}&       1.013\sym{***}&       1.012\sym{***}\\
                    &   (0.00229)         &   (0.00229)         &   (0.00229)         &   (0.00229)         &   (0.00229)         &   (0.00229)         &   (0.00229)         &   (0.00229)         &   (0.00229)         &   (0.00229)         \\
\hline
Field fe            & $\checkmark$        & $\checkmark$        & $\checkmark$        & $\checkmark$        & $\checkmark$        & $\checkmark$        & $\checkmark$        & $\checkmark$        & $\checkmark$        & $\checkmark$        \\
Year fe             & $\checkmark$        & $\checkmark$        & $\checkmark$        & $\checkmark$        & $\checkmark$        & $\checkmark$        & $\checkmark$        & $\checkmark$        & $\checkmark$        & $\checkmark$        \\
\(N\)               &      334124         &      334124         &      334124         &      334124         &      334124         &      334124         &      334124         &      334124         &      334124         &      334124         \\
\textit{BIC}        &   2562441.2         &   2562234.1         &   2562300.7         &   2562435.2         &   2562032.7         &   2561523.0         &   2561882.7         &   2561152.3         &   2561166.1         &   2560728.6         \\
\bottomrule
\multicolumn{5}{l}{\footnotesize Standard errors in parentheses}\\
\multicolumn{5}{l}{\footnotesize \sym{*} \(p<0.05\), \sym{**} \(p<0.01\), \sym{***} \(p<0.001\)}\\
\end{tabular}
\end{table*}
\end{landscape}

\input{appendix}


\begin{thebibliography}{68}
\expandafter\ifx\csname natexlab\endcsname\relax\def\natexlab#1{#1}\fi
\providecommand{\url}[1]{\texttt{#1}}
\providecommand{\href}[2]{#2}
\providecommand{\path}[1]{#1}
\providecommand{\DOIprefix}{doi:}
\providecommand{\ArXivprefix}{arXiv:}
\providecommand{\URLprefix}{URL: }
\providecommand{\Pubmedprefix}{pmid:}
\providecommand{\doi}[1]{\href{http://dx.doi.org/#1}{\path{#1}}}
\providecommand{\Pubmed}[1]{\href{pmid:#1}{\path{#1}}}
\providecommand{\bibinfo}[2]{#2}
\ifx\xfnm\relax \def\xfnm[#1]{\unskip,\space#1}\fi
\bibitem[{Belenzon and Schankerman(2013)}]{Belenzon-spreading-2013}
\bibinfo{author}{Belenzon, S.}, \bibinfo{author}{Schankerman, M.},
  \bibinfo{year}{2013}.
\newblock \bibinfo{title}{Spreading the word: Geography, policy, and knowledge
  spillovers}.
\newblock \bibinfo{journal}{The Review of Economics and Statistics}
  \bibinfo{volume}{95}, \bibinfo{pages}{884--903}.
\newblock \DOIprefix\doi{10.1162/REST_a_00334}.
\bibitem[{Biancani et~al.(2014)Biancani, McFarland and
  Dahlander}]{Biancani2014semiformal}
\bibinfo{author}{Biancani, S.}, \bibinfo{author}{McFarland, D.A.},
  \bibinfo{author}{Dahlander, L.}, \bibinfo{year}{2014}.
\newblock \bibinfo{title}{The semiformal organization}.
\newblock \bibinfo{journal}{Organization Science} \bibinfo{volume}{25},
  \bibinfo{pages}{1306--1324}.
\newblock \DOIprefix\doi{10.1287/orsc.2013.0882}.
\bibitem[{Bikard(2018)}]{Bikard-made-2018}
\bibinfo{author}{Bikard, M.}, \bibinfo{year}{2018}.
\newblock \bibinfo{title}{Made in academia: The effect of institutional origin
  on inventors' attention to science}.
\newblock \bibinfo{journal}{Organization Science} \bibinfo{volume}{29},
  \bibinfo{pages}{818--836}.
\newblock \DOIprefix\doi{10.1287/orsc.2018.1206}.
\bibitem[{Bikard and Marx(2019)}]{Bikard-bridging-2019}
\bibinfo{author}{Bikard, M.}, \bibinfo{author}{Marx, M.}, \bibinfo{year}{2019}.
\newblock \bibinfo{title}{Bridging academia and industry: How geographic hubs
  connect university science and corporate technology}.
\newblock \bibinfo{journal}{Management Science} \bibinfo{volume}{0},
  \bibinfo{pages}{0}.
\newblock \DOIprefix\doi{10.1287/mnsc.2019.3385}.
\bibitem[{Bordons et~al.(1999)Bordons, Zulueta, Romero and
  Barrig{\'o}n}]{bordons1999measuring}
\bibinfo{author}{Bordons, M.}, \bibinfo{author}{Zulueta, M.},
  \bibinfo{author}{Romero, F.}, \bibinfo{author}{Barrig{\'o}n, S.},
  \bibinfo{year}{1999}.
\newblock \bibinfo{title}{Measuring interdisciplinary collaboration within a
  university: The effects of the multidisciplinary research programme}.
\newblock \bibinfo{journal}{Scientometrics} \bibinfo{volume}{46},
  \bibinfo{pages}{383--398}.
\newblock \DOIprefix\doi{10.1007/BF02459599}.
\bibitem[{Bromham et~al.(2016)Bromham, Dinnage and
  Hua}]{Bromham-interdisc-2016}
\bibinfo{author}{Bromham, L.}, \bibinfo{author}{Dinnage, R.},
  \bibinfo{author}{Hua, X.}, \bibinfo{year}{2016}.
\newblock \bibinfo{title}{Interdisciplinary research has consistently lower
  funding success}.
\newblock \bibinfo{journal}{Nature} \bibinfo{volume}{534},
  \bibinfo{pages}{684--687}.
\newblock \DOIprefix\doi{10.1038/nature18315}.
\bibitem[{Bruce et~al.(2004)Bruce, Lyall, Tait and Williams}]{Bruce2004idr}
\bibinfo{author}{Bruce, A.}, \bibinfo{author}{Lyall, C.},
  \bibinfo{author}{Tait, J.}, \bibinfo{author}{Williams, R.},
  \bibinfo{year}{2004}.
\newblock \bibinfo{title}{Interdisciplinary integration in {Europe}: the case
  of the {Fifth Framework} programme}.
\newblock \bibinfo{journal}{Futures} \bibinfo{volume}{36},
  \bibinfo{pages}{457--470}.
\newblock \DOIprefix\doi{10.1016/j.futures.2003.10.003}.
\bibitem[{Carayol and Thi(2005)}]{Carayol-why-2005}
\bibinfo{author}{Carayol, N.}, \bibinfo{author}{Thi, T.U.N.},
  \bibinfo{year}{2005}.
\newblock \bibinfo{title}{Why do academic scientists engage in
  interdisciplinary research?}
\newblock \bibinfo{journal}{Research Evaluation} \bibinfo{volume}{14},
  \bibinfo{pages}{70--79}.
\newblock \DOIprefix\doi{10.3152/147154405781776355}.
\bibitem[{Cassi et~al.(2017)Cassi, Champeimont, Mescheba and
  de~Turckheim}]{Cassi-analysing-2017}
\bibinfo{author}{Cassi, L.}, \bibinfo{author}{Champeimont, R.},
  \bibinfo{author}{Mescheba, W.}, \bibinfo{author}{de~Turckheim, E.},
  \bibinfo{year}{2017}.
\newblock \bibinfo{title}{Analysing institutions interdisciplinarity by
  extensive use of rao-stirling diversity index}.
\newblock \bibinfo{journal}{PLOS ONE} \bibinfo{volume}{12},
  \bibinfo{pages}{e0170296}.
\newblock \DOIprefix\doi{10.1371/journal.pone.0170296}.
\bibitem[{Cech and Rubin(2004)}]{Cech-nurturing-2004}
\bibinfo{author}{Cech, T.R.}, \bibinfo{author}{Rubin, G.M.},
  \bibinfo{year}{2004}.
\newblock \bibinfo{title}{Nurturing interdisciplinary research}.
\newblock \bibinfo{journal}{Nature Structural \& Molecular Biology}
  \bibinfo{volume}{11}, \bibinfo{pages}{1166--1169}.
\newblock \DOIprefix\doi{10.1038/nsmb1204-1166}.
\bibitem[{Cockburn and Henderson(1996)}]{Cockburn-public-1996}
\bibinfo{author}{Cockburn, I.}, \bibinfo{author}{Henderson, R.},
  \bibinfo{year}{1996}.
\newblock \bibinfo{title}{Public-private interaction in pharmaceutical
  research}.
\newblock \bibinfo{journal}{Proceedings of the National Academy of Sciences}
  \bibinfo{volume}{93}, \bibinfo{pages}{12725--12730}.
\newblock \DOIprefix\doi{10.1073/pnas.93.23.12725}.
\bibitem[{Debackere and Veugelers(2005)}]{Debackere-role-2005}
\bibinfo{author}{Debackere, K.}, \bibinfo{author}{Veugelers, R.},
  \bibinfo{year}{2005}.
\newblock \bibinfo{title}{The role of academic technology transfer
  organizations in improving industry science links}.
\newblock \bibinfo{journal}{Research Policy} \bibinfo{volume}{34},
  \bibinfo{pages}{321--342}.
\newblock \DOIprefix\doi{10.1016/j.respol.2004.12.003}.
\bibitem[{D'Este et~al.(2019)D'Este, Llopis, Rentocchini and
  Yegros}]{Este-relation-2019}
\bibinfo{author}{D'Este, P.}, \bibinfo{author}{Llopis, O.},
  \bibinfo{author}{Rentocchini, F.}, \bibinfo{author}{Yegros, A.},
  \bibinfo{year}{2019}.
\newblock \bibinfo{title}{The relationship between interdisciplinarity and
  distinct modes of university-industry interaction}.
\newblock \bibinfo{journal}{Research Policy} \bibinfo{volume}{48},
  \bibinfo{pages}{103799}.
\newblock \DOIprefix\doi{10.1016/j.respol.2019.05.008}.
\bibitem[{Feller(2006)}]{feller2006multiple}
\bibinfo{author}{Feller, I.}, \bibinfo{year}{2006}.
\newblock \bibinfo{title}{Multiple actors, multiple settings, multiple
  criteria: issues in assessing interdisciplinary research}.
\newblock \bibinfo{journal}{Research Evaluation} \bibinfo{volume}{15},
  \bibinfo{pages}{5--15}.
\newblock \DOIprefix\doi{10.3152/147154406781776020}.
\bibitem[{Fieller(1954)}]{fieller1954some}
\bibinfo{author}{Fieller, E.C.}, \bibinfo{year}{1954}.
\newblock \bibinfo{title}{Some problems in interval estimation}.
\newblock \bibinfo{journal}{Journal of the Royal Statistical Society: Series B
  (Methodological)} \bibinfo{volume}{16}, \bibinfo{pages}{175--185}.
\newblock \DOIprefix\doi{10.1111/j.2517-6161.1954.tb00159.x}.
\bibitem[{Fleming(2001)}]{Lee2001recombinant}
\bibinfo{author}{Fleming, L.}, \bibinfo{year}{2001}.
\newblock \bibinfo{title}{Recombinant uncertainty in technological search}.
\newblock \bibinfo{journal}{Management Science} \bibinfo{volume}{47},
  \bibinfo{pages}{117--132}.
\newblock \DOIprefix\doi{10.1287/mnsc.47.1.117.10671}.
\bibitem[{Fleming and Sorenson(2004)}]{Fleming-science-2004}
\bibinfo{author}{Fleming, L.}, \bibinfo{author}{Sorenson, O.},
  \bibinfo{year}{2004}.
\newblock \bibinfo{title}{Science as a map in technological search}.
\newblock \bibinfo{journal}{Strategic Management Journal} \bibinfo{volume}{25},
  \bibinfo{pages}{909--928}.
\newblock \DOIprefix\doi{10.1002/smj.384}.
\bibitem[{Fontana et~al.(2022)Fontana, Iori, Sciabolazza and
  Souza}]{fontana2022interdisciplinarity}
\bibinfo{author}{Fontana, M.}, \bibinfo{author}{Iori, M.},
  \bibinfo{author}{Sciabolazza, V.L.}, \bibinfo{author}{Souza, D.},
  \bibinfo{year}{2022}.
\newblock \bibinfo{title}{The interdisciplinarity dilemma: public versus
  private interests}.
\newblock \bibinfo{journal}{Research Policy} \bibinfo{volume}{51},
  \bibinfo{pages}{104553}.
\newblock \DOIprefix\doi{10.1016/j.respol.2022.104553}.
\bibitem[{Gambardella(1992)}]{Gambardella-competitive-1992}
\bibinfo{author}{Gambardella, A.}, \bibinfo{year}{1992}.
\newblock \bibinfo{title}{Competitive advantages from in-house scientific
  research: The us pharmaceutical industry in the 1980s}.
\newblock \bibinfo{journal}{Research Policy} \bibinfo{volume}{21},
  \bibinfo{pages}{391--407}.
\newblock \DOIprefix\doi{10.1016/0048-7333(92)90001-K}.
\bibitem[{Gates et~al.(2019)Gates, Ke, Varol and
  Barab\'asi}]{Gates-nature-2019}
\bibinfo{author}{Gates, A.J.}, \bibinfo{author}{Ke, Q.},
  \bibinfo{author}{Varol, O.}, \bibinfo{author}{Barab\'asi, A.L.},
  \bibinfo{year}{2019}.
\newblock \bibinfo{title}{\emph{Nature}'s reach: narrow work has broad impact}.
\newblock \bibinfo{journal}{Nature} \bibinfo{volume}{575},
  \bibinfo{pages}{32--34}.
\newblock \DOIprefix\doi{10.1038/d41586-019-03308-7}.
\bibitem[{Giuliani et~al.(2010)Giuliani, Morrison, Pietrobelli and
  Rabellotti}]{Giuliani2010who}
\bibinfo{author}{Giuliani, E.}, \bibinfo{author}{Morrison, A.},
  \bibinfo{author}{Pietrobelli, C.}, \bibinfo{author}{Rabellotti, R.},
  \bibinfo{year}{2010}.
\newblock \bibinfo{title}{Who are the researchers that are collaborating with
  industry? an analysis of the wine sectors in chile, south africa and italy}.
\newblock \bibinfo{journal}{Research Policy} \bibinfo{volume}{39},
  \bibinfo{pages}{748--761}.
\newblock \DOIprefix\doi{10.1016/j.respol.2010.03.007}.
\bibitem[{Gregorio and Shane(2003)}]{Gregorio-why-2003}
\bibinfo{author}{Gregorio, D.D.}, \bibinfo{author}{Shane, S.},
  \bibinfo{year}{2003}.
\newblock \bibinfo{title}{Why do some universities generate more start-ups than
  others?}
\newblock \bibinfo{journal}{Research Policy} \bibinfo{volume}{32},
  \bibinfo{pages}{209--227}.
\newblock \DOIprefix\doi{10.1016/S0048-7333(02)00097-5}.
\bibitem[{Haans et~al.(2016)Haans, Pieters and He}]{haans2016thinking}
\bibinfo{author}{Haans, R.F.}, \bibinfo{author}{Pieters, C.},
  \bibinfo{author}{He, Z.L.}, \bibinfo{year}{2016}.
\newblock \bibinfo{title}{Thinking about u: Theorizing and testing u-and
  inverted u-shaped relationships in strategy research}.
\newblock \bibinfo{journal}{Strategic Management Journal} \bibinfo{volume}{37},
  \bibinfo{pages}{1177--1195}.
\newblock \DOIprefix\doi{10.1002/smj.2399}.
\bibitem[{Hackett et~al.(2021)Hackett, Leahey, Parker, Rafols, Hampton, Corte,
  Chavarro, Drake, Penders, Sheble, Vermeulen and
  Vision}]{hackett2021synthesis}
\bibinfo{author}{Hackett, E.J.}, \bibinfo{author}{Leahey, E.},
  \bibinfo{author}{Parker, J.N.}, \bibinfo{author}{Rafols, I.},
  \bibinfo{author}{Hampton, S.E.}, \bibinfo{author}{Corte, U.},
  \bibinfo{author}{Chavarro, D.}, \bibinfo{author}{Drake, J.M.},
  \bibinfo{author}{Penders, B.}, \bibinfo{author}{Sheble, L.},
  \bibinfo{author}{Vermeulen, N.}, \bibinfo{author}{Vision, T.J.},
  \bibinfo{year}{2021}.
\newblock \bibinfo{title}{Do synthesis centers synthesize? a semantic analysis
  of topical diversity in research}.
\newblock \bibinfo{journal}{Research Policy} \bibinfo{volume}{50},
  \bibinfo{pages}{104069}.
\newblock \DOIprefix\doi{10.1016/j.respol.2020.104069}.
\bibitem[{Heinze et~al.(2009)Heinze, Shapira, Rogers and
  Senker}]{Heinze-2009-org}
\bibinfo{author}{Heinze, T.}, \bibinfo{author}{Shapira, P.},
  \bibinfo{author}{Rogers, J.D.}, \bibinfo{author}{Senker, J.M.},
  \bibinfo{year}{2009}.
\newblock \bibinfo{title}{Organizational and institutional influences on
  creativity in scientific research}.
\newblock \bibinfo{journal}{Research Policy} \bibinfo{volume}{38},
  \bibinfo{pages}{610--623}.
\newblock \DOIprefix\doi{10.1016/j.respol.2009.01.014}.
\bibitem[{Jacobs and Frickel(2009)}]{Jacobs-2009-idr}
\bibinfo{author}{Jacobs, J.A.}, \bibinfo{author}{Frickel, S.},
  \bibinfo{year}{2009}.
\newblock \bibinfo{title}{Interdisciplinarity: A critical assessment}.
\newblock \bibinfo{journal}{Annual Review of Sociology} \bibinfo{volume}{35},
  \bibinfo{pages}{43--65}.
\newblock \DOIprefix\doi{10.1146/annurev-soc-070308-115954}.
\bibitem[{Ke(2018)}]{Ke-compare-2018}
\bibinfo{author}{Ke, Q.}, \bibinfo{year}{2018}.
\newblock \bibinfo{title}{Comparing scientific and technological impact of
  biomedical research}.
\newblock \bibinfo{journal}{Journal of Informetrics} \bibinfo{volume}{12},
  \bibinfo{pages}{706--717}.
\newblock \DOIprefix\doi{10.1016/j.joi.2018.06.010}.
\bibitem[{Ke(2020a)}]{Ke-analysis-2020}
\bibinfo{author}{Ke, Q.}, \bibinfo{year}{2020}a.
\newblock \bibinfo{title}{An analysis of the evolution of science-technology
  linkage in biomedicine}.
\newblock \bibinfo{journal}{Journal of Informetrics} \bibinfo{volume}{14},
  \bibinfo{pages}{101074}.
\newblock \DOIprefix\doi{10.1016/j.joi.2020.101074}.
\bibitem[{Ke(2020b)}]{Ke-tech-2020}
\bibinfo{author}{Ke, Q.}, \bibinfo{year}{2020}b.
\newblock \bibinfo{title}{Technological impact of biomedical research: The role
  of basicness and novelty}.
\newblock \bibinfo{journal}{Research Policy} \bibinfo{volume}{49},
  \bibinfo{pages}{104071}.
\newblock \DOIprefix\doi{10.1016/j.respol.2020.104071}.
\bibitem[{Ke et~al.(2017)Ke, Ahn and Sugimoto}]{Ke2017twitter}
\bibinfo{author}{Ke, Q.}, \bibinfo{author}{Ahn, Y.Y.},
  \bibinfo{author}{Sugimoto, C.R.}, \bibinfo{year}{2017}.
\newblock \bibinfo{title}{A systematic identification and analysis of
  scientists on twitter}.
\newblock \bibinfo{journal}{PLOS ONE} \bibinfo{volume}{12},
  \bibinfo{pages}{e0175368}.
\newblock \DOIprefix\doi{10.1371/journal.pone.0175368}.
\bibitem[{Klein(2008)}]{Klein2008}
\bibinfo{author}{Klein, J.T.}, \bibinfo{year}{2008}.
\newblock \bibinfo{title}{Evaluation of interdisciplinary and transdisciplinary
  research: A literature review}.
\newblock \bibinfo{journal}{American Journal of Preventive Medicine}
  \bibinfo{volume}{35}, \bibinfo{pages}{S116--S123}.
\newblock \DOIprefix\doi{10.1016/j.amepre.2008.05.010}.
\bibitem[{Larivière and Gingras(2010)}]{Lariviere-relation-2010}
\bibinfo{author}{Larivière, V.}, \bibinfo{author}{Gingras, Y.},
  \bibinfo{year}{2010}.
\newblock \bibinfo{title}{On the relationship between interdisciplinarity and
  scientific impact}.
\newblock \bibinfo{journal}{Journal of the American Society for Information
  Science and Technology} \bibinfo{volume}{61}, \bibinfo{pages}{126--131}.
\newblock \DOIprefix\doi{10.1002/asi.21226}.
\bibitem[{Leahey et~al.(2017)Leahey, Beckman and
  Stanko}]{Leahey-prominent-2017}
\bibinfo{author}{Leahey, E.}, \bibinfo{author}{Beckman, C.M.},
  \bibinfo{author}{Stanko, T.L.}, \bibinfo{year}{2017}.
\newblock \bibinfo{title}{Prominent but less productive: The impact of
  interdisciplinarity on scientists’ research}.
\newblock \bibinfo{journal}{Administrative Science Quarterly}
  \bibinfo{volume}{62}, \bibinfo{pages}{105--139}.
\newblock \DOIprefix\doi{10.1177/0001839216665364}.
\bibitem[{Ledford(2015)}]{Ledford-how-2015}
\bibinfo{author}{Ledford, H.}, \bibinfo{year}{2015}.
\newblock \bibinfo{title}{How to solve the world's biggest problems}.
\newblock \bibinfo{journal}{Nature} \bibinfo{volume}{525},
  \bibinfo{pages}{308--311}.
\newblock \DOIprefix\doi{10.1038/525308a}.
\bibitem[{Levitt and Thelwall(2008)}]{Levitt-multi-2008}
\bibinfo{author}{Levitt, J.M.}, \bibinfo{author}{Thelwall, M.},
  \bibinfo{year}{2008}.
\newblock \bibinfo{title}{Is multidisciplinary research more highly cited? a
  macrolevel study}.
\newblock \bibinfo{journal}{Journal of the American Society for Information
  Science and Technology} \bibinfo{volume}{59}, \bibinfo{pages}{1973--1984}.
\newblock \DOIprefix\doi{10.1002/asi.20914}.
\bibitem[{Leydesdorff et~al.(2019)Leydesdorff, Wagner and
  Bornmann}]{Leydesdorff-idr-2019}
\bibinfo{author}{Leydesdorff, L.}, \bibinfo{author}{Wagner, C.S.},
  \bibinfo{author}{Bornmann, L.}, \bibinfo{year}{2019}.
\newblock \bibinfo{title}{Interdisciplinarity as diversity in citation patterns
  among journals: Rao-stirling diversity, relative variety, and the gini
  coefficient}.
\newblock \bibinfo{journal}{Journal of Informetrics} \bibinfo{volume}{13},
  \bibinfo{pages}{255--269}.
\newblock \DOIprefix\doi{10.1016/j.joi.2018.12.006}.
\bibitem[{Lind and Mehlum(2010)}]{lind2010or}
\bibinfo{author}{Lind, J.T.}, \bibinfo{author}{Mehlum, H.},
  \bibinfo{year}{2010}.
\newblock \bibinfo{title}{With or without u? the appropriate test for a
  u-shaped relationship}.
\newblock \bibinfo{journal}{Oxford Bulletin of Economics and Statistics}
  \bibinfo{volume}{72}, \bibinfo{pages}{109--118}.
\newblock \DOIprefix\doi{10.1111/j.1468-0084.2009.00569.x}.
\bibitem[{Lowe and Phillipson(2006)}]{Lowe-reflexive-2006}
\bibinfo{author}{Lowe, P.}, \bibinfo{author}{Phillipson, J.},
  \bibinfo{year}{2006}.
\newblock \bibinfo{title}{Reflexive interdisciplinary research: The making of a
  research programme on the rural economy and land use}.
\newblock \bibinfo{journal}{Journal of Agricultural Economics}
  \bibinfo{volume}{57}, \bibinfo{pages}{165--184}.
\newblock \DOIprefix\doi{10.1111/j.1477-9552.2006.00045.x}.
\bibitem[{Metzger and Zare(1999)}]{Metzger-1999-idr}
\bibinfo{author}{Metzger, N.}, \bibinfo{author}{Zare, R.N.},
  \bibinfo{year}{1999}.
\newblock \bibinfo{title}{Interdisciplinary research: From belief to reality}.
\newblock \bibinfo{journal}{Science} \bibinfo{volume}{283},
  \bibinfo{pages}{642--643}.
\newblock \DOIprefix\doi{10.1126/science.283.5402.642}.
\bibitem[{Meyer(2000)}]{Meyer-does-2000}
\bibinfo{author}{Meyer, M.}, \bibinfo{year}{2000}.
\newblock \bibinfo{title}{Does science push technology? patents citing
  scientific literature}.
\newblock \bibinfo{journal}{Research Policy} \bibinfo{volume}{29},
  \bibinfo{pages}{409--434}.
\newblock \DOIprefix\doi{10.1016/S0048-7333(99)00040-2}.
\bibitem[{Millar(2013)}]{Millar2013idr}
\bibinfo{author}{Millar, M.M.}, \bibinfo{year}{2013}.
\newblock \bibinfo{title}{Interdisciplinary research and the early career: The
  effect of interdisciplinary dissertation research on career placement and
  publication productivity of doctoral graduates in the sciences}.
\newblock \bibinfo{journal}{Research Policy} \bibinfo{volume}{42},
  \bibinfo{pages}{1152--1164}.
\newblock \DOIprefix\doi{10.1016/j.respol.2013.02.004}.
\bibitem[{Misra et~al.(2009)Misra, Harvey, Stokols, Pine, Fuqua, Shokair and
  Whiteley}]{Misra2009evaluating}
\bibinfo{author}{Misra, S.}, \bibinfo{author}{Harvey, R.H.},
  \bibinfo{author}{Stokols, D.}, \bibinfo{author}{Pine, K.H.},
  \bibinfo{author}{Fuqua, J.}, \bibinfo{author}{Shokair, S.M.},
  \bibinfo{author}{Whiteley, J.M.}, \bibinfo{year}{2009}.
\newblock \bibinfo{title}{Evaluating an interdisciplinary undergraduate
  training program in health promotion research}.
\newblock \bibinfo{journal}{American Journal of Preventive Medicine}
  \bibinfo{volume}{36}, \bibinfo{pages}{358--365}.
\newblock \DOIprefix\doi{10.1016/j.amepre.2008.11.014}.
\bibitem[{Molas-Gallart et~al.(2014)Molas-Gallart, Rafols and
  Tang}]{Gallart-relation-2014}
\bibinfo{author}{Molas-Gallart, J.}, \bibinfo{author}{Rafols, I.},
  \bibinfo{author}{Tang, P.}, \bibinfo{year}{2014}.
\newblock \bibinfo{title}{On the relationship between interdisciplinarity and
  impact: Different modalities of interdisciplinarity lead to different types
  of impact}.
\newblock \bibinfo{journal}{Journal of Science Policy and Research Management}
  \bibinfo{volume}{29}, \bibinfo{pages}{69--89}.
\newblock \DOIprefix\doi{10.20801/jsrpim.29.2_3_69}.
\bibitem[{Narin et~al.(1997)Narin, Hamilton and Olivastro}]{Narin-linkage-1997}
\bibinfo{author}{Narin, F.}, \bibinfo{author}{Hamilton, K.S.},
  \bibinfo{author}{Olivastro, D.}, \bibinfo{year}{1997}.
\newblock \bibinfo{title}{The increasing linkage between {U.S.} technology and
  public science}.
\newblock \bibinfo{journal}{Research Policy} \bibinfo{volume}{26},
  \bibinfo{pages}{317--330}.
\newblock \DOIprefix\doi{10.1016/S0048-7333(97)00013-9}.
\bibitem[{Narin and Olivastro(1992)}]{Narin-status-1992}
\bibinfo{author}{Narin, F.}, \bibinfo{author}{Olivastro, D.},
  \bibinfo{year}{1992}.
\newblock \bibinfo{title}{Status report: Linkage between technology and
  science}.
\newblock \bibinfo{journal}{Research Policy} \bibinfo{volume}{21},
  \bibinfo{pages}{237--249}.
\newblock \DOIprefix\doi{10.1016/0048-7333(92)90018-Y}.
\bibitem[{Narin and Olivastro(1998)}]{Narin1998linkage}
\bibinfo{author}{Narin, F.}, \bibinfo{author}{Olivastro, D.},
  \bibinfo{year}{1998}.
\newblock \bibinfo{title}{Linkage between patents and papers: An interim epo/us
  comparison}.
\newblock \bibinfo{journal}{Scientometrics} \bibinfo{volume}{41},
  \bibinfo{pages}{51--59}.
\newblock \DOIprefix\doi{10.1007/BF02457966}.
\bibitem[{{National Academy of Sciences} et~al.(2005){National Academy of
  Sciences}, {National Academy of Engineering} and {Institute of
  Medicine}}]{NAP11153}
\bibinfo{author}{{National Academy of Sciences}}, \bibinfo{author}{{National
  Academy of Engineering}}, \bibinfo{author}{{Institute of Medicine}},
  \bibinfo{year}{2005}.
\newblock \bibinfo{title}{Facilitating Interdisciplinary Research}.
\newblock \bibinfo{publisher}{The National Academies Press},
  \bibinfo{address}{Washington, DC}.
\newblock \DOIprefix\doi{10.17226/11153}.
\bibitem[{Porter et~al.(2007)Porter, Cohen, Roessner and
  Perreault}]{porter2007measuring}
\bibinfo{author}{Porter, A.L.}, \bibinfo{author}{Cohen, A.S.},
  \bibinfo{author}{Roessner, J.D.}, \bibinfo{author}{Perreault, M.},
  \bibinfo{year}{2007}.
\newblock \bibinfo{title}{Measuring researcher interdisciplinarity}.
\newblock \bibinfo{journal}{Scientometrics} \bibinfo{volume}{72},
  \bibinfo{pages}{117--147}.
\newblock \DOIprefix\doi{10.1007/s11192-007-1700-5}.
\bibitem[{Porter and Rafols(2009)}]{Porter-science-2009}
\bibinfo{author}{Porter, A.L.}, \bibinfo{author}{Rafols, I.},
  \bibinfo{year}{2009}.
\newblock \bibinfo{title}{Is science becoming more interdisciplinary? measuring
  and mapping six research fields over time}.
\newblock \bibinfo{journal}{Scientometrics} \bibinfo{volume}{81},
  \bibinfo{pages}{719}.
\newblock \DOIprefix\doi{10.1007/s11192-008-2197-2}.
\bibitem[{Porter et~al.(2006)Porter, Roessner, Cohen and
  Perreault}]{Porter2006idr}
\bibinfo{author}{Porter, A.L.}, \bibinfo{author}{Roessner, J.D.},
  \bibinfo{author}{Cohen, A.S.}, \bibinfo{author}{Perreault, M.},
  \bibinfo{year}{2006}.
\newblock \bibinfo{title}{Interdisciplinary research: meaning, metrics and
  nurture}.
\newblock \bibinfo{journal}{Research Evaluation} \bibinfo{volume}{15},
  \bibinfo{pages}{187--195}.
\newblock \DOIprefix\doi{10.3152/147154406781775841}.
\bibitem[{Porter and Rossini(1985)}]{Porter-peer-1985}
\bibinfo{author}{Porter, A.L.}, \bibinfo{author}{Rossini, F.A.},
  \bibinfo{year}{1985}.
\newblock \bibinfo{title}{Peer review of interdisciplinary research proposals}.
\newblock \bibinfo{journal}{Science, Technology, \& Human Values}
  \bibinfo{volume}{10}, \bibinfo{pages}{33--38}.
\newblock \DOIprefix\doi{10.1177/016224398501000304}.
\bibitem[{Rafols et~al.(2012)Rafols, Leydesdorff, O'Hare, Nightingale and
  Stirling}]{Rafols-how-2012}
\bibinfo{author}{Rafols, I.}, \bibinfo{author}{Leydesdorff, L.},
  \bibinfo{author}{O'Hare, A.}, \bibinfo{author}{Nightingale, P.},
  \bibinfo{author}{Stirling, A.}, \bibinfo{year}{2012}.
\newblock \bibinfo{title}{How journal rankings can suppress interdisciplinary
  research: A comparison between innovation studies and business \&
  management}.
\newblock \bibinfo{journal}{Research Policy} \bibinfo{volume}{41},
  \bibinfo{pages}{1262--1282}.
\newblock \DOIprefix\doi{10.1016/j.respol.2012.03.015}.
\bibitem[{Rinia et~al.(2001)Rinia, van Leeuwen, van Vuren and van
  Raan}]{Rinia-influence-2001}
\bibinfo{author}{Rinia, E.J.}, \bibinfo{author}{van Leeuwen, T.N.},
  \bibinfo{author}{van Vuren, H.G.}, \bibinfo{author}{van Raan, A.F.J.},
  \bibinfo{year}{2001}.
\newblock \bibinfo{title}{Influence of interdisciplinarity on peer-review and
  bibliometric evaluations in physics research}.
\newblock \bibinfo{journal}{Research Policy} \bibinfo{volume}{30},
  \bibinfo{pages}{357--361}.
\newblock \DOIprefix\doi{10.1016/S0048-7333(00)00082-2}.
\bibitem[{Roach and Cohen(2013)}]{Roach-lens-2013}
\bibinfo{author}{Roach, M.}, \bibinfo{author}{Cohen, W.M.},
  \bibinfo{year}{2013}.
\newblock \bibinfo{title}{Lens or prism? patent citations as a measure of
  knowledge flows from public research}.
\newblock \bibinfo{journal}{Management Science} \bibinfo{volume}{59},
  \bibinfo{pages}{504--525}.
\newblock \DOIprefix\doi{10.1287/mnsc.1120.1644}.
\bibitem[{Silva et~al.(2013)Silva, Rodrigues, Oliveira and
  da~F.~Costa}]{Silva-quantify-2013}
\bibinfo{author}{Silva, F.}, \bibinfo{author}{Rodrigues, F.},
  \bibinfo{author}{Oliveira, O.}, \bibinfo{author}{da~F.~Costa, L.},
  \bibinfo{year}{2013}.
\newblock \bibinfo{title}{Quantifying the interdisciplinarity of scientific
  journals and fields}.
\newblock \bibinfo{journal}{Journal of Informetrics} \bibinfo{volume}{7},
  \bibinfo{pages}{469--477}.
\newblock \DOIprefix\doi{10.1016/j.joi.2013.01.007}.
\bibitem[{Steele and Stier(2000)}]{Steele-impact-2000}
\bibinfo{author}{Steele, T.W.}, \bibinfo{author}{Stier, J.C.},
  \bibinfo{year}{2000}.
\newblock \bibinfo{title}{The impact of interdisciplinary research in the
  environmental sciences: a forestry case study}.
\newblock \bibinfo{journal}{Journal of the American Society for Information
  Science} \bibinfo{volume}{51}, \bibinfo{pages}{476--484}.
\newblock
  \DOIprefix\doi{10.1002/(SICI)1097-4571(2000)51:5<476::AID-ASI8>3.0.CO;2-G}.
\bibitem[{Stirling(2007)}]{stirling2007general}
\bibinfo{author}{Stirling, A.}, \bibinfo{year}{2007}.
\newblock \bibinfo{title}{A general framework for analysing diversity in
  science, technology and society}.
\newblock \bibinfo{journal}{Journal of the Royal Society Interface}
  \bibinfo{volume}{4}, \bibinfo{pages}{707--719}.
\newblock \DOIprefix\doi{10.1098/rsif.2007.0213}.
\bibitem[{Sun et~al.(2021)Sun, Livan, Ma and Latora}]{sun2021interdisciplinary}
\bibinfo{author}{Sun, Y.}, \bibinfo{author}{Livan, G.}, \bibinfo{author}{Ma,
  A.}, \bibinfo{author}{Latora, V.}, \bibinfo{year}{2021}.
\newblock \bibinfo{title}{Interdisciplinary researchers attain better long-term
  funding performance}.
\newblock \bibinfo{journal}{Communications Physics} \bibinfo{volume}{4},
  \bibinfo{pages}{263}.
\newblock \DOIprefix\doi{10.1038/s42005-021-00769-z}.
\bibitem[{Szostak(2008)}]{szostak2008classification}
\bibinfo{author}{Szostak, R.}, \bibinfo{year}{2008}.
\newblock \bibinfo{title}{Classification, interdisciplinarity, and the study of
  science}.
\newblock \bibinfo{journal}{Journal of Documentation} \bibinfo{volume}{64},
  \bibinfo{pages}{319--332}.
\newblock \DOIprefix\doi{10.1108/00220410810867551}.
\bibitem[{Tussen et~al.(2000)Tussen, Buter and
  Van~Leeuwen}]{tussen2000technological}
\bibinfo{author}{Tussen, R.J.W.}, \bibinfo{author}{Buter, R.K.},
  \bibinfo{author}{Van~Leeuwen, T.N.}, \bibinfo{year}{2000}.
\newblock \bibinfo{title}{Technological relevance of science: An assessment of
  citation linkages between patents and research papers}.
\newblock \bibinfo{journal}{Scientometrics} \bibinfo{volume}{47},
  \bibinfo{pages}{389--412}.
\newblock \DOIprefix\doi{10.1023/A:1005603513439}.
\bibitem[{Uzzi et~al.(2013)Uzzi, Mukherjee, Stringer and
  Jones}]{Uzzi-atypical-2013}
\bibinfo{author}{Uzzi, B.}, \bibinfo{author}{Mukherjee, S.},
  \bibinfo{author}{Stringer, M.}, \bibinfo{author}{Jones, B.},
  \bibinfo{year}{2013}.
\newblock \bibinfo{title}{Atypical combinations and scientific impact}.
\newblock \bibinfo{journal}{Science} \bibinfo{volume}{342},
  \bibinfo{pages}{468--472}.
\newblock \DOIprefix\doi{10.1126/science.1240474}.
\bibitem[{{van Rijnsoever} et~al.(2008){van Rijnsoever}, Hessels and
  Vandeberg}]{Rijnsoever2008resource}
\bibinfo{author}{{van Rijnsoever}, F.J.}, \bibinfo{author}{Hessels, L.K.},
  \bibinfo{author}{Vandeberg, R.L.}, \bibinfo{year}{2008}.
\newblock \bibinfo{title}{A resource-based view on the interactions of
  university researchers}.
\newblock \bibinfo{journal}{Research Policy} \bibinfo{volume}{37},
  \bibinfo{pages}{1255--1266}.
\newblock \DOIprefix\doi{10.1016/j.respol.2008.04.020}.
\bibitem[{Verhoeven et~al.(2016)Verhoeven, Bakker and
  Veugelers}]{Verhoeven-measuring-2016}
\bibinfo{author}{Verhoeven, D.}, \bibinfo{author}{Bakker, J.},
  \bibinfo{author}{Veugelers, R.}, \bibinfo{year}{2016}.
\newblock \bibinfo{title}{Measuring technological novelty with patent-based
  indicators}.
\newblock \bibinfo{journal}{Research Policy} \bibinfo{volume}{45},
  \bibinfo{pages}{707--723}.
\newblock \DOIprefix\doi{10.1016/j.respol.2015.11.010}.
\bibitem[{Wagner et~al.(2011)Wagner, Roessner, Bobb, Klein, Boyack, Keyton,
  Rafols and B\"orner}]{Wagner-approache-2011}
\bibinfo{author}{Wagner, C.S.}, \bibinfo{author}{Roessner, J.D.},
  \bibinfo{author}{Bobb, K.}, \bibinfo{author}{Klein, J.T.},
  \bibinfo{author}{Boyack, K.W.}, \bibinfo{author}{Keyton, J.},
  \bibinfo{author}{Rafols, I.}, \bibinfo{author}{B\"orner, K.},
  \bibinfo{year}{2011}.
\newblock \bibinfo{title}{Approaches to understanding and measuring
  interdisciplinary scientific research (idr): A review of the literature}.
\newblock \bibinfo{journal}{Journal of Informetrics} \bibinfo{volume}{5},
  \bibinfo{pages}{14--26}.
\newblock \DOIprefix\doi{10.1016/j.joi.2010.06.004}.
\bibitem[{Wang et~al.(2015)Wang, Thijs and Gl\"{a}nzel}]{Wang-impact-2015}
\bibinfo{author}{Wang, J.}, \bibinfo{author}{Thijs, B.},
  \bibinfo{author}{Gl\"{a}nzel, W.}, \bibinfo{year}{2015}.
\newblock \bibinfo{title}{Interdisciplinarity and impact: Distinct effects of
  variety, balance, and disparity}.
\newblock \bibinfo{journal}{PLOS ONE} \bibinfo{volume}{10},
  \bibinfo{pages}{e0127298}.
\newblock \DOIprefix\doi{10.1371/journal.pone.0127298}.
\bibitem[{Wang and Verberne(2021)}]{wang2021two}
\bibinfo{author}{Wang, J.}, \bibinfo{author}{Verberne, S.},
  \bibinfo{year}{2021}.
\newblock \bibinfo{title}{Two tales of science technology linkage: Patent
  in-text versus front-page references}.
\newblock \bibinfo{journal}{arXiv preprint arXiv:2103.08931} .
\bibitem[{Wang and Wiborg~Schneider(2020)}]{Wang-consistency-2020}
\bibinfo{author}{Wang, Q.}, \bibinfo{author}{Wiborg~Schneider, J.},
  \bibinfo{year}{2020}.
\newblock \bibinfo{title}{Consistency and validity of interdisciplinarity
  measures}.
\newblock \bibinfo{journal}{Quantitative Science Studies} \bibinfo{volume}{1},
  \bibinfo{pages}{239--263}.
\newblock \DOIprefix\doi{10.1162/qss_a_00011}.
\bibitem[{Yegros-Yegros et~al.(2015)Yegros-Yegros, Rafols and
  D’Este}]{Yegros-interdisc-2015}
\bibinfo{author}{Yegros-Yegros, A.}, \bibinfo{author}{Rafols, I.},
  \bibinfo{author}{D’Este, P.}, \bibinfo{year}{2015}.
\newblock \bibinfo{title}{Does interdisciplinary research lead to higher
  citation impact? the different effect of proximal and distal
  interdisciplinarity}.
\newblock \bibinfo{journal}{PLOS ONE} \bibinfo{volume}{10},
  \bibinfo{pages}{e0135095}.
\newblock \DOIprefix\doi{10.1371/journal.pone.0135095}.

\end{thebibliography}
\end{document}